\documentclass[11pt,journal]{IEEEtran}

\onecolumn

\newcommand{\subparagraph}{}
\usepackage{nccmath}
\usepackage{mathtools}
\IEEEoverridecommandlockouts

\usepackage{graphicx} 
\usepackage{epsfig} 

\usepackage{amsmath}
\usepackage{amsthm}
\newtheorem*{assumption*}{\assumptionnumber}
\providecommand{\assumptionnumber}{}
\makeatletter
\newenvironment{assumption}[1]
 {%
  \renewcommand{\assumptionnumber}{Assumption #1}%
  \begin{assumption*}%
  \protected@edef\@currentlabel{#1}%
 }
 {%
  \end{assumption*}
 }
\makeatother

\usepackage{amssymb}
\usepackage{tikz}
\usepackage[export]{adjustbox}
\usepackage{mathcomSTEP}
\usepackage[utf8]{inputenc}
\usepackage{enumerate}
\usepackage[outdir=./]{epstopdf}
\usepackage{pgfplots,setspace}
\pgfplotsset{compat=1.14}
\usetikzlibrary{positioning}
\usepackage[hidelinks]{hyperref}
\newcommand{\mmse}{\mathsf {mmse}}
\newcommand{\ind}{\mathsf {ind}}

\newcommand{\CE}{\mathsf {CE}}
\newcommand{\ML}{\mathsf {ML}}

\newcommand{\TC}{\mathsf {TC}}

\newcommand{\tr}{\mathrm {Tr}}

\newcommand{\PSD}{\mathbb S_{+}^n}
\newcommand{\Enc}{\mathrm {Enc}}

\newcommand{\enc}{\mathsf {enc}}

\newcommand{\reals}{\mathbb R}

\newcommand{\ex}[1]{\ensuremath{\mathbb{E}\left[ #1\right]}}
\newcommand{\simiid}{\overset{\mathrm{iid}}{\sim}}

\tikzstyle{int}=[draw, fill=blue!10, minimum height = 1cm, minimum width=1.5cm,thick ]
\tikzstyle{sint}=[draw, fill=blue!10, minimum height = 0.5cm, minimum width=0.8cm,thick ]
\tikzstyle{sum}=[circle, fill=blue!10, draw=black,line width=1pt,minimum size = 0.5cm, thick ]
\tikzstyle{ssum}=[circle, fill=blue!10,draw=black,line width=1pt,minimum size = 0.1cm]
\tikzstyle{int1}=[draw, fill=blue!10, minimum height = 0.5cm, minimum width=1cm,thick ]
\tikzstyle{enc}=[draw, fill=blue!10, minimum height = 2.7cm, minimum width=1cm,thick ]
\tikzstyle{int}=[draw, fill=blue!10, minimum height = 1cm, minimum width=1.5cm,thick ]

\author{
\IEEEauthorblockN{Alon Kipnis\IEEEauthorrefmark{1}, Stefano Rini\IEEEauthorrefmark{2} and Andrea J. Goldsmith\IEEEauthorrefmark{3} \\}

\IEEEauthorblockA{%
\IEEEauthorrefmark{1}
Department of Statistics, Stanford University, Stanford, CA 94305 USA\\
}

\IEEEauthorblockA{%
\IEEEauthorrefmark{2}
Department of Electrical Engineering, National Chiao-Tung University, Hsinchu, 30010, Taiwan \\
}

\IEEEauthorblockA{%
\IEEEauthorrefmark{3}
Department of Electrical Engineering, Stanford University, Stanford, CA 94305 USA\\
}

\thanks{This paper was presented in part at the IEEE International Symposium on Information Theory (ISIT), 2016 \cite{kipnis2016multiterminal} and 2017 \cite{KipnisRini2017ISIT}, and at the Information Theory Workshop (ITW), 2015 \cite{KipnisRini2015}.}
\thanks{
This work was supported in part by the Koret Foundation, the NSF Center for Science of Information (CSoI) under grant CCF-0939370, the NSF grants DMS-1418362 and DMS-1407813, and the NSF-BSF under grant 1609695. }
}

\title{
The Rate-Distortion Risk in\\Estimation from Compressed Data
}

\begin{document}

\graphicspath{{./Figs/}}
\maketitle

\thispagestyle{plain}
\pagestyle{plain}

\begin{abstract}
Consider the problem of estimating a latent signal from a lossy compressed version of the data when the compressor is agnostic to the relation between the signal and the data.
This situation arises in a host of modern applications when data is transmitted or stored prior to determining the downstream inference task. 
Given a bitrate constraint and a distortion measure between the data and its compressed version, let us consider the joint distribution achieving Shannon's rate-distortion (RD) function. Given an estimator and a loss function associated with the downstream inference task, define the rate-distortion risk as the expected loss under the RD-achieving distribution. 
We provide general conditions under which the operational risk in estimating from the compressed data is asymptotically equivalent to the RD risk.
The main theoretical tools to prove this equivalence are transportation-cost inequalities in conjunction with properties of compression codes achieving Shannon's RD function.
Whenever such equivalence holds, a recipe for designing estimators from datasets undergoing lossy compression without specifying the actual compression technique emerges: design the estimator to minimize the RD risk. 
Our conditions simplified in the special cases of discrete memoryless or multivariate normal data. For these scenarios, we derive explicit expressions for the RD risk of several estimators and compare them to the optimal source coding performance associated with full knowledge of the relation between the latent signal and the data.
 \end{abstract}

\begin{IEEEkeywords}
Source coding; 
Compression for estimation;
Remote source coding;
Mismatched source coding;
Compress and estimate;
\end{IEEEkeywords}


\section{Introduction}
Digital systems suffer from intrinsic limitations in the number of bits they can store, communicate, or process. 
As the acquired real-world data is compressed in a lossy manner before any inference takes place, the performance of inference procedures are intrinsically dictated by the quality of the compression applied to the raw data and, in particular, by the number of bits available for its compressed representation. \par
In this paper, we study the effect of lossy compression on the performance of estimators from datasets undergoing data compression. Specifically, we characterize the performance of estimation from the compressed data using an expression that is independent of the specific compression technique; it only depends on the distribution of the data and the compression bitrate, the number of bits-per-symbol (bps) in the compressed representation. 
The benefit of our characterization is twofold: it provides (i) a way to evaluate the risk in estimation from datasets undergoing lossy compression and (ii) a way to design such estimators by optimizing their performance with respect to the resulting risk.  \par
\begin{figure}
\begin{center}
\begin{tikzpicture}[node distance=2cm,auto,>=latex]
\node at (0,0) (th) {$\theta$};
\node [int1, right of = th, node distance = 2cm](pxy){$P_{Y^n|{\theta}}$};

\node [right of = pxy, node distance = 2cm](data){$Y^n$};
\node [int1, below of = data, node distance = 1.5cm, align = center](comp) {\small lossy \\ \small compression};
%
\draw[<-] (comp) -- +(-1.5,0.5) node[left,xshift=0.5cm,align = center,text width=4cm] {\small $R$ (bitrate constraint)};

\draw[<-] (comp) -- +(-1.5,-0.5) node[left,xshift=0.4cm, align = center,text width=5cm] {\small $d(y^n,z^n)$ (distortion measure) };

\node [left] (dest) [below of = th, node distance = 3 cm]{$\hat{\theta}$};

\node[int1, below of = pxy, node distance = 3cm] (est)  {\small estimator $\phi$};
\node[below of=comp, node distance = 1.5 cm] (z) {$Z^n$};
\draw[->,line width=1pt] (th) -- (pxy);
\draw[->,line width=1pt] (pxy) -- (data);
\draw[->,line width=1pt] (data) -- (comp);
\draw[->,line width=1pt] (comp) -- (z);
\draw[->,line width=1pt] (z) -- (est);
\draw[->,line width=1pt] (est) -- (dest);

\draw[<-, line width=0.5pt, dotted] (data) -- +(0.7,+0.7) node[above right,align=center] {\scriptsize{data} };

\draw[<-, line width=0.5pt, dotted] (th) -- +(-0.7,+0.7) node[above left,align=center] {\scriptsize{signal} };

\draw[<-, line width=0.5pt, dotted] (z) -- +(0.65,-0.65) node[below right,align=center] {\scriptsize{compressed} \\ \scriptsize{representation} };

\draw[<-, line width=0.5pt, dotted] (dest) -- +(-0.65,-0.65) node[below left,align=center] {\scriptsize{estimated} \\ \scriptsize{signal} };

\end{tikzpicture}
\caption{
A conceptual representation of 
signal estimation from compressed data. The goal is to recover the latent signal $\theta$ from a lossy compressed version $Z^n$ of the data $Y^n$ at bitrate $R$, while the compression is according to a distortion measure between $Y^n$ and $Z^n$. 
}
\label{fig:system_intro}
\end{center}
\end{figure}
To better describe our motivation and results, consider the conceptual representation provided in
Figure~\ref{fig:system_intro}. 
The latent signal is represented by $\theta$ and the observed data is the $n$-length sequence $Y^n$. We compress $Y^n$ in a lossy manner at bitrate $R$, i.e., using $nR$ bits, and denote the resulting compressed representation by $Z^n$. 
%
Finally, $Z^n$ is utilized to produce an estimate $\hat{\theta} = \phi(Z^n)$ of $\theta$. The performance of the system is measured in terms of the expected loss \begin{align}
    D_\phi \equiv 
    \ex{\ell(\theta, \hat{\theta})}
    \label{eq:true_risk},
\end{align}
for some given loss function $\ell(\theta,\hat{\theta})$. 
%
In this paper, we characterize $D_\phi$ in a special version of the setting in Figure~\ref{fig:system_intro}. Conceptually speaking, we assume the existence of a distortion measure $d(y^n,z^n)$, and suppose that the expectation over $d(Y^n,Z^n)$ approaches its information-theoretic minimum subject to the bitrate-$R$ compression constraint. Namely, the lossy compression operation attains Shannon's distortion-rate (DR) function of $Y^n$ at bitrate $R$. Compression codes possessing this property are denoted \emph{good} rate-distortion (RD) codes. 
As we demonstrate, combining known properties of good RD codes \cite{KostinaVerdu} with classical results in transportation theory \cite{villani2008optimal} leads to conditions for the convergence of $D_\phi$ to an expression that is independent of the specific compression code. This expression,
denoted as the \emph{rate-distortion risk} of $\phi$, is the expected loss when $\phi$ applied to $Z^{*n}$ instead of $Z^n$ of \eqref{eq:true_risk}, where the joint distribution of $Y^n$ and $Z^{*n}$ attains the DR function of $Y^n$ with respect to $d(y^n,z^n)$. 
\subsection*{Estimate-and-Compress vs. Compress-and-Estimate}
In the study of the compression and estimation system of Figure~\ref{fig:system_intro}, particular attention is naturally given to the optimal trade-off between bitrate $R$ and expected loss \eqref{eq:true_risk}. It is well-known that this optimal trade-off is attained by an \emph{estimate-and-compress} (EC) strategy \cite{DobrushinTsybakov,WolfZiv1970, berger1971rate}. Namely, first, estimate $\theta$ from the data $Y^n$, then compress this estimate according to the available bitrate $R$. For example, when the pair $(\theta, Y^n)$ defines an ergodic information source, the optimal trade-off between bitrate and expected loss is attained when the estimation step uses the Bayes estimator for $\theta$ and the compression step approaches the Shannon RD function of the output of the Bayes estimator. The resulting optimal trade-off is described by the \emph{indirect} DR function or its inverse, the indirect RD function \cite{berger1971rate,witsenhausen1980indirect}. Note that, the EC strategy generally depends on the model $P_{Y^n|\theta}$ and the loss $\ell$, hence EC is infeasible when the estimator depends on the yet-to-be-known downstream application or when the same dataset is used for more than one application \cite{adams2015big,ma2015remote}. 
Arguably, in the examples above, the compression operation may be designed to only account for the distortion between $Y^n$ and $Z^n$ (e.g., Euclidean distance, Hamming distance, or empirical mutual information), while the model $P_{Y^n|\theta}$ and the estimation procedures are determined post-compression. The resulting compression and estimation scenario is known as \emph{compress-and-estimate} (CE) \cite{Schizas2008,KipnisWiener2019}. 
\par
Despite the apparent prevalence of systems implementing a CE strategy, its performance characterization has so far been limited to very specific models $P_{Y^n|\theta}$ \cite{Schizas2008,KipnisWiener2019}, compression procedures $Y^n\to Z^n$ \cite{lapidoth1997role,KipnisReeves2020}, or estimators $\phi$
\cite{zhu2014quantized,donoho2002kolmogorov}. The goal of this paper is to characterize the performance of CE under general conditions. \par
Henceforth, we refer to $D_\phi$ of \eqref{eq:true_risk} as the \emph{CE risk} of the estimator $\phi$. This risk depends on the specific compression code $Y^n \to Z^n$. On the other hand, the RD risk of $\phi$, our candidate for describing the CE risk, only depends on the bitrate and the distortion measure $d(y^n,z^n)$. Consequently, a by-product of the equivalence of these two risks is the asymptotic independence of the CE risk in the specific compressor employed, as long as this compressor attains the RD function of $Y^n$.
\par
To better highlight the differences between the EC and CE scenarios, as well as to motivate our characterization of the latter, we present an illustrative example employing the familiar setting of Gaussian signals and quadratic loss functions. 

\subsection*{Example: Compressed Gaussian Measurements
}
Suppose that the data $Y^n$ consists of $n$ independent measurements from a bivariate normal distribution $\Ncal(0,\Sigma_y)$, for some known covariance matrix $\Sigma_y$. We compress $Y^n$ to an $nR$-bit representation $Z^n$. The minimal MSE in recovering $Y^n$ from $Z^n$ is Shannon's quadratic DR function at bitrate $R$, 
denoted as $D_y(R)$. In our case, $D_y(R)$ is obtained by waterfilling over the two eigenvalues of $\Sigma_y$ \cite{berger1971rate}. 
Let us assume that $Z^n$ is a version of $Y^n$ obtained using a good RD code at bitrate $R$, i.e, we have $\ex{\|Y^n-Z^n\|_2^2}/n \to D_y(R)$ as $n\to \infty$. Next, consider the following model
\begin{align}
\label{eq:2d_Gaussian_model}
\begin{pmatrix}
Y_{i,1} \\
Y_{i,2} 
\end{pmatrix}
 = \sqrt{\frac{\gamma}{2}} \theta_i \begin{pmatrix}
1 \\
1
\end{pmatrix} + \begin{pmatrix}
W_{i,1} \\
W_{i,2}
\end{pmatrix}, \quad i=1,\ldots,n,
\end{align}
where $\theta_i$, $W_{i,1}$, and $W_{i,2}$ are iid standard normal random variables and $\gamma>0$ is a fixed signal-to-noise (SNR) parameter.
%
%
%
Two major challenges arise in estimating $\theta=(\theta_1,\ldots,\theta_n)$ from $Z^n$: choosing the estimator $\phi$ and characterizing the resulting risk. One natural choice for $\phi$ is the maximum likelihood (ML) estimator of $\theta$ from $Y^n$, given as
\[
\phi^{\ML}(y_{i,1},y_{i,2}) =
\frac{1}{\sqrt{2\gamma}}\left( y_{i,1} + y_{i,2} \right). 
\]
An attractive property of $\phi^{\ML}$ is its independence of the compression scheme and even of the bitrate of this compression. Provided some conditions are met, our main result implies that the quadratic risk of $\phi^{\ML}$ when applied to $Z^n$ converges to the so-called RD risk of $\phi^{\ML}$:
the MSE in estimating $\theta$ by applying $\phi^{\ML}$ to $Z^{*n}$, where $P_{Y^nZ^{*n}}$ attains the quadratic Gaussian RD function of $\Ncal(0,\Sigma_y)$ at the bitrate $R$. Figure~\ref{fig:Gaussian_intro} illustrates this RD risk based on a closed-form expression we provide in Section~\ref{sec:Gaussian} below. 
\par
%
The asymptotic equivalence between the CE and the RD risks suggests that we can attain MSE lower than the RD risk of $\phi^\ML$ by utilizing the Bayes estimator of $\theta$ from $Z^{*n}$. 
We denote the resulting risk as the \emph{Bayes RD} risk; it is given by
\begin{align}
    \label{eq:2d_Gaussian_CE}
D^*(R) = \frac{1}{1+\gamma} +  \frac{\gamma }{1+\gamma}  2^{[R-R_{\gamma}]^+-2R},
\end{align}
where $R_{\gamma} \equiv \frac{1}{2}\log(1+\gamma)$ and 
$[x]^+ \equiv \max\{x,0\}$. Under some conditions, we show that the Bayes RD risk is the minimal MSE in estimating $\theta$ from \emph{any} $nR$-bitrate representation of $Y^n$ attaining $D_y(R)$. In this sense, \eqref{eq:2d_Gaussian_CE} provides the minimal MSE in the CE scenario.  
\par
As it turns out, \eqref{eq:2d_Gaussian_CE} can be dramatically different from the optimal performance in this compression and estimation setting. To derive the latter, we use the EC strategy: form the Bayes estimator of $\theta$ from $Y^n$
\begin{align}
    \label{eq:U}
\ex{\theta_i | Y_{1,i},Y_{2,i}} = \frac{\sqrt{\gamma/2} }{1+\gamma}(Y_{1,i}+Y_{2,i}),\quad i=1,\ldots,n,
\end{align}
then compress $\{\ex{\theta_i | Y_{1,i},Y_{2,i}}\}_{i=1}^n$ using $nR$ bits in an optimal manner subject to a MSE criterion. The resulting MSE in this procedure approaches the indirect DR function of $\theta$, given as
\begin{align}
    \label{eq:2d_Gaussian_indirect}
D_{\ind}(R) = \frac{1}{1+\gamma} + \frac{\gamma}{1+\gamma}2^{-2R}.
\end{align}
Figure~\ref{fig:Gaussian_intro} compares the three distortion functions discussed above for a specific choice of the SNR parameter $\gamma$. $D^*(R)$ of 
\eqref{eq:2d_Gaussian_CE} and $D_{\ind}(R)$ of \eqref{eq:2d_Gaussian_indirect} coincide at small compression rates $R\leq R_{\gamma}$, but converge to their infinite bitrate asymptotic $1/(1+\gamma)$ at different rates.  Intuitively, the reason for this difference is that, under CE, the compressor $Y^n \to Z^n$ does not distinguish between signal and noise components, as this distinction is only possible when the model \eqref{eq:2d_Gaussian_model} is known before compression. 
\begin{figure}
\begin{center}
\begin{tikzpicture}
\def\SNR{3}
\def\MM{2}
\def\Rmax{5}

    \begin{axis}[
    width=8cm,
    height=6cm,
    legend style={at={(1,1)},
      anchor=north east, legend columns=1},
    ylabel={MSE},
    xlabel={$R$ [BPS]},
    ytick={0,0.25,0.5,1},
    yticklabels={0,$\frac{1}{1+\gamma}$,,1},
    xtick={1,2,3,4},
    xticklabels={1,2,3,4},
    legend cell align={left},
    ymin=0,
    xmin=0,
    xmax=\Rmax,
    ymax=1.05,
    ]

\addplot[,color=orange, mark = *, mark options={scale=0.2}, style=thick] 
coordinates {(0,2) (1,2)};

\addplot[color=red, style=thick] 
coordinates {(0,2) (1,2)};

\addplot[color=blue, mark=|, mark options={scale=0.5}, style=thick] 
coordinates {(0,2) (1,2)};

\addplot[dashed, color=black, style=thick] coordinates {(0,2) (1,2)};

\addlegendentry{\scriptsize{$D^*_{\phi^{\ML}}$ (RD risk of ML estimator)}};
\addlegendentry{\scriptsize{$D^*$ (RD risk of Bayes estimator)}};
\addlegendentry{\scriptsize{$D_{\ind}$} (optimal compression) };

\node[xshift=1cm] at (axis cs: 1,0.6) {\scriptsize $R_0 = \frac{1}{2}\log(1+\gamma)$};

\addplot[domain=0:0.5*ln(1+\SNR)/ln(2), color=orange, samples=7, style=thick, mark=*, mark options={scale=0.2}] 
({x},{
(1+2^(-2*x)*(\SNR-1))/\SNR });

\addplot[domain=0.5*ln(1+\SNR)/ln(2):\Rmax, color=orange, samples=17, style=thick, mark=*, mark options={scale=0.2}] 
({x},{
2^(-(4/\MM)*(x +(\MM-1)*ln(1+ \SNR)/2/ln(2)))
+((1-2^(-2*x))*(1-2^(-2*x) + 2^(-(2/\MM)*(x-ln(1+ \SNR)/2/ln(2))) ) )/ \SNR 
  });

\addplot[domain=0:\Rmax, color=blue, samples=27, style=thick, mark=|, mark options={scale=0.5}] 
({x},{1/(1+\SNR) + \SNR/(1 + \SNR)*2^(-2*x)});

\addplot[domain=0:0.5*ln(1+\SNR)/ln(2), color=red, samples=7, style=thick] 
({x},{1/(1+\SNR) + 2^(-2*x)*\SNR   /(1+\SNR) });

\addplot[domain=0.5*ln(1+\SNR)/ln(2):\Rmax, color=red, samples=17, style=thick] 
({x},{
1/(1+\SNR) + \SNR/(1+\SNR) * 2^(-(2*x/\MM + (\MM-1)/\MM * ln(1+\SNR)/ln(2))) });

\addplot[domain=0:\Rmax, color=black, samples=27, style=thick, dotted] 
({x},{1/(1+\SNR)});

\addplot[domain=(\Rmax-1):\Rmax, color=black, samples=27, style=thick, dotted] 
({x},{1/\SNR});

\node[] (one_over_g) at (axis cs: \Rmax,1/\SNR) {};
\node[] (Rcrit) at (axis cs: 1.5,0.6) {};


\addplot[dotted, dotted] 
coordinates {(0.5*ln(1+\SNR)/ln(2),0)  (0.5*ln(1+\SNR)/ln(2),0.55)};

\end{axis}
\node[right] at (one_over_g) {\scriptsize $\frac{1}{\gamma}$};        
\end{tikzpicture}
\caption{
Mean-squared error in estimating a Gaussian signal in the bivariate normal measurements model \eqref{eq:2d_Gaussian_model} when compressing these measurements at bitrate $R$. The SNR parameter $\gamma$ is $3$. The indirect DR function $D_{\ind}(R)$ describes the optimal performance, attained by estimate-and-compress. The performance under compress-and-estimate using a quadratic distortion for compressing the measurements and the maximum likelihood estimator or the Bayes estimator for estimating the latent signal, are described by the RD risks of these estimators. 
}
\label{fig:Gaussian_intro}
\end{center}
\end{figure}
By establishing the equivalence between the CE and the RD risks, we can interpret the difference $D^*(R) - D_{\ind}(R)$ as the price of compressing the data without knowing the model \eqref{eq:2d_Gaussian_model} that is ultimately used in inference. 
Such lack of knowledge is the norm rather than the exception in modern data science and signal acquisition applications, hence the importance of establishing the connection between the RD risk and the risk in the CE scenario. \par
We may also consider a scenario in which the model \eqref{eq:2d_Gaussian_model} is initially assumed, the EC strategy for the task of estimating $\theta$ is utilized, but a new downstream task is declared: recover the original data $Y^n$. In this case, we apply the CE analysis with  \[
U_i \equiv \ex{\theta_i | Y_{1,i},Y_{2,i}},\quad i=1,\ldots,n,
\]
as the data and $Y^n$ as the latent signal. In Appendix~\ref{app:proof_intro}, we show that the difference between the Bayes RD risk $D_{U}^*(R)$ in this case and the optimal compression performance, given here by the DR function of $Y^n$, is
\begin{align}
    \label{eq:2d_diff}
    D_{U}^*(R)- D_y(R)=\left( 1 - 2^{-[R-R_{\gamma}]^+} \right)^2.
\end{align}
Applying a similar interpretation of the RD risk as before, \eqref{eq:2d_diff} is the price of compressing according to the wrong downstream task. This price is zero for $R\leq R_{\gamma}$, but can be significant at high bitrates and SNR.
\par
Interestingly, the situation is considerably different in the univariate counterpart of \eqref{eq:2d_Gaussian_model}, i.e.,
\[
Y_i = \sqrt{\gamma}\theta_i + W_i,\quad i=1,\ldots,n. 
\]
In this case, the Bayes RD risk and the indirect DR function coincide for all values of $R$, i.e., no difference in performance between the CE and the EC scenarios in this case. This different behavior of the univariate and multivariate cases is perhaps the reason why the difference between CE and EC strategies was so far not given considerable attention. 
\subsection*{Related Works}
The problem of estimation from compressed data has a long history in source coding. When the pair $(\theta, Y_n)$ defines a stationary ergodic information source, the problem of encoding $Y^n$ for estimating $\theta$ is known as the \emph{indirect} (other names are \emph{remote} or \emph{noisy}) source coding problem 
\cite[Ch 3.5]{berger1971rate}, \cite{DobrushinTsybakov,WolfZiv1970, witsenhausen1980indirect, kostina2016nonasymptotic}. The minimum of all achievable distortions using a compression code of rate not exceeding $R$ is denoted as the
indirect DR function. When $\theta$ is a vector of parameters, the problem of compressing $Y^n$ for recovering $\theta$ was studied under the names  \emph{compression for estimation} \cite{han1987hypothesis,720540} or \emph{task-specific compression} \cite{1093858,1093935, GrayConference2006,shlezinger2019hardware}. These problems, and in particular their multi-terminal versions, have received much recent interest due to their relevance in machine learning \cite{duchi2014optimality,  Kipnis2019MeanEF,barnes2019learning}.\par
The CE setting differs from these works since the compression is assumed optimal only with respect to the data $Y^n$ rather than the estimator of $\theta$.
The works of 
 \cite{Ishwar2005,Wang2018RobustDC} study a multi-terminal version of this problem, and are motivated by the robustness in performance due to encoders oblivious to other system components. 
To our knowledge, the term CE distortion first appeared in \cite{Schizas2008} and is inspired by the compress-and-forward coding scheme for the relay channel \cite{cover1979capacity}. The works of \cite{RiniKipnisTWC, Schizas2008} analyze this distortion in a multi-terminal setup for a class of vector Gaussian sources. Other settings with exact CE characterization include compressing the samples of the Wiener process \cite{KipnisWiener2019} and compressing the measurements obtained via a sequence of random linear projections \cite{kipnis2018isit}. 
Recently, the work of \cite{KipnisReeves2020} provided conditions under which the risk of a Lipschitz estimator from data compressed using a random spherical code equals the risk of the same estimator when applied to a corrupted version of the data by an additive white Gaussian noise (AWGN). In particular, under this form of compression, the risk \eqref{eq:true_risk} of any Lipschitz estimator converges to its 
RD risk whenever the RD-achieving distribution $P_{Z^{* n}Y^n}$ is Gaussian. In this paper, we generalize this result from \cite{KipnisReeves2020} to a broad class of RD-achieving distributions and loss functions. Specifically, our setup applies to any situation where the RD-achieving distribution satisfies a suitable transportation-cost inequality \cite{raginsky2013concentration, villani2008optimal}. As we explain below, this condition holds quite broadly in the important cases of memoryless data or when the RD-achieving distribution is Gaussian. 
\par
%
%
%
One may also view the CE scheme as an instance of the mismatched source coding problems considered in \cite{lapidoth1997role} and \cite{kontoyiannis2006mismatched}. Indeed, as pointed out in \cite{berger1971rate, witsenhausen1980indirect,dembo2003minimax}, in some situations, we can attain the indirect DR function by compressing the data according to a distortion measure 
\begin{equation}
\tilde{d}(y^n,z^n) \equiv \inf_{\phi} \ex{\ell(\theta, \phi(z^n)) |Y^n=y^n}. \label{eq:amended}
\end{equation}
Namely, when the compression attains the DR function of $Y^n$ with respect to $\tilde{d}=\tilde{d}(y^n,z^n)$, the optimal source coding performance for estimating $\theta$ are also attained. The said mismatch can possibly arise when $Y^n$ is compressed so as to minimize some $d \neq \tilde{d}$ (note that $\tilde{d}$ in \eqref{eq:amended} requires the model $\theta\to Y^n$, so compressing with respect to $\tilde{d}$ is not a valid CE scheme).\par
Finally, Donoho \cite{donoho2000counting,donoho2002kolmogorov} considered the properties of optimal compressors in the model
\begin{equation}
    \label{eq:Gaussian_sequence}
Y_i = \theta_i + \sigma W_i,\quad W_i \simiid \Ncal(0,1),
\end{equation}
as a universal denoiser (estimator) for the vector of means 
$\theta = (\theta_1,\ldots,\theta_n)$. He noted that a good RD code for compressing $Y^n$ at the quadratic distortion level $\sigma^2$ induces a distribution on $Z^n$ that approaches the true posterior $P_{\theta_1|Y_1}$ (although this posterior is never revealed to the compressor). Consequently, the resulting expected quadratic loss in estimating $\theta$ is $2\ex{\var[\theta_1|Y_1]}$, the risk in sampling from this posterior. Translated to our terminology, the phenomena above says that, under \eqref{eq:Gaussian_sequence}, the CE risk with respect to the quadratic loss ($\ell = \|\cdot\|_2^2/n$) of the maximum likelihood estimator $\phi^{\ML}(x) = x$ at coding rate $R = R(\sigma^2) = \frac{1}{2} \log\left(\frac{\var[Y_1]}{\sigma^2}\right)$ and distortion $d = \|\cdot\|_2^2$, converges to 
\[
 D_{\phi^{\ML}}^*(R,\theta) = 2\ex{\var\left[\theta_1|Y_1\right]}.
\]
Unlike \cite{donoho2000counting,donoho2002kolmogorov}, the compression in our case is a limitation imposed on the system rather than a feature yielding an estimator if employed at the particular bitrate $R(\sigma^2)$. In particular, we consider the compression of $Y^n$ at arbitrary bitrates or distortion levels, and not only at the level
$\sigma^2$ considered by Donoho. 


\subsection*{Contributions}
%
In this paper, we derive conditions for the convergence of the risk \eqref{eq:true_risk} of an estimator $\phi$ to its RD risk and evaluate this RD risk in several special cases. 
The first result of this paper, as described in Section~\ref{sec:main}, relates to cases where the forward RD-achieving distribution $P_{Z^{* n}|Y^n}$ is unique and satisfying a transportation-cost inequality with a constant that is inversely-linear in the problem dimension $n$.
In this case, we obtain a strong bound on the difference between the RD risk of any Lipschitz estimator and its CE risk when fed with the output of any compression code $P_{Z^n|Y^n}$ of bitrate $R_n$ satisfying the target distortion $D_y$.
%
In particular, this bound vanishes for good RD codes satisfying $R_n\to R(D_y)$ and whose output distribution $P_{Z^n}$ converges to $P_{Z^{*n}}$ in the normalized relative entropy. 
Next, we analyze the TC condition in two important cases: (1) $P_{Y^n Z^{*n}}$ is a product measure over a discrete domain (Section~\ref{sec:discrete}), or (2) $P_{Y^n Z^{* n}}$ is a Gaussian measure on $\reals^n$ (Section~\ref{sec:Gaussian}). 
To demonstrate the usefulness of our results and highlight the difference between the CE scenario and the optimal performance, we derive closed-form expressions for the RD risk in several useful cases: (i) When $\theta$ is an iid binary signal, $Y^n$ is its noisy observation under a binary symmetric channel, and $\phi$ is the Bayes estimator. 
(ii) When $\theta$ is a Gaussian iid signal and each $Y_i$ is an $m$-dimensional vector representing a sequence of its noisy observations. In the scenarios above we derive the RD risk of both the Bayes estimator and the maximum likelihood estimator of $\theta$ from the uncompressed data. 
In all cases, we compare the resulting risk to the optimal performance described by the indirect DR function. 

\subsection*{Paper Organization}
The remainder of the paper is organized as follows: Section \ref{sec:Problem Formulation}
formalizes the problem and introduce notation to describe the results. In Section~\ref{sec:TC} we review relevant topics in transportation theory and provide a general characterization of the difference between the CE and RD risks.
In Sections~\ref{sec:discrete} and \ref{sec:Gaussian}, we specialize to the settings where the data is discrete and Gaussian, respectively. Concluding remarks are provided in Section~\ref{sec:Conclusions}.

\subsection*{Notation}
We use capital and calligraphic letters, respectively, to denote the random variable $U$ over an alphabet $\Ucal$. Therefore, unless stated otherwise, we implicitly assume a probability space and a topology on $\Ucal$ such that $U$ is measurable. We use the notation $X^n \simiid P_X$ to indicate that $X^n$ is a random sequence obtained by sampling independently $n$ times from the distribution $P_X$.  
We denote an $n$-length sequence over $\Ucal$ as $u^n = (u_1,\ldots,u_n) \in \Ucal^n$.
In some specific case when $\Ucal = \reals^m$, we denote $\uv^n = (\uv_1,\ldots,\uv_n) \in \reals^{m \times n}$. Matrices are also denoted by bold capital letters, hence, depending on the context, $\Uv$ may denote either a random vector over $\reals^m$ or a matrix. The identity matrix of size $n\times n$ is denoted by $\Id_n$. The set of $n\times n$ positive definite matrices is denoted by $\PSD$. The operator norm of a matrix $\Sigma$ is denoted by $\|\Sigma\|$. We denote by  $\lambda_1(\Sigma),\ldots,\lambda_n(\Sigma)$ the eigenvalues of the matrix $\Sigma$, ordered by magnitude. The notation $\diag\left[\lambda_1,\ldots,\lambda_n \right]$ indicates a diagonal matrix with diagonal entries $\lambda_1,\ldots,\lambda_n$.
The indicator function of the set $A$ is denoted $\mathbf 1_A$. For $x\in \reals$, we define $[x]^+ \equiv \max\{x , 0\}$ and $\log^+(x) \equiv [\log(x)]^+$, $x>0$. For $x^n\in \reals^n$, $\|x^n\|_2$ denotes its Euclidean norm. The symbol $\ll$ indicates the absolute continuity relation between measures. The basis of all logarithms is $2$. 
%

\section{Problem Formulation}
\label{sec:Problem Formulation}
\begin{figure}
\centering
\begin{tikzpicture}[node distance=2cm,auto,>=latex]
\node at (0,0) (th) {$\theta$};
\node[above = 0cm of th] {$(\Theta,\rho)$};
\node [int1, right of = th, node distance = 1.5cm](pxy){$P_{Y^n|{\theta}}$};
\node [right of = pxy, node distance = 1.5cm](data){$Y^n$};
\node[above = 0cm of data] {$\Ycal^n$};
\node [int1,right of = data, node distance = 1.5cm](enc){$\Enc$};
\node [int1] (dec) [below of= enc, node distance = 2 cm] {$\rm Dec$};

\node (z) [below of= data, node distance = 2 cm] {$Z^n$};

\node[below of = z, node distance = 0.5 cm] {$(\Zcal^n, \nu)$};

\node [int1] (est) [below of= pxy, node distance = 2 cm] {$\phi$};

\node [right] (dest) [below of=th, node distance = 2 cm]{$\hat{\theta}$};
\node[below = 0cm of dest] {$(\Theta,\rho)$};

\draw[->,line width=1pt] (th) -- (pxy);
\draw[->,line width=1pt] (pxy) -- (data);
\draw[->,line width=1pt] (data) -- (enc);
\draw[->,line width=1pt] (enc) -- node[right, xshift=-0cm, yshift = 0cm] {\small $R_n$}  (dec);

\draw[->,line width=1pt] (dec) -- (z);

\draw[->,line width=1pt] (z) -- (est);

\draw[->,line width=1pt] (est) -- (dest);

\draw[<->, line width=0.5pt, dashed] (data) -- node {$d$}  (z);
\draw[<->, line width=0.5pt, dashed] (th) -- node {$\rho$} (dest);

\end{tikzpicture}

\caption{Compression and estimation system model. 
$\theta$ and  $Y^n$ represent the signal and the data, respectively. $Z^n$ is obtained by compressing the data using $nR_n$ bits. $\phi$ is an estimator of $\theta$ from $Z^n$. We assume that $(\Theta,\rho)$ and $(\Zcal^n, \nu)$ are metric spaces and while $d$ is a distortion measure on $\Ycal^n \times \Zcal^n$. 
}
\label{fig:system}
\end{figure}

Consider the lossy compression and estimation problem illustrated in Figure~\ref{fig:system}. This figure is a detailed version of Figure~\ref{fig:system_intro} where the lossy compression operation is separated into its encoding and decoding parts. 
The data is represented by the $n$-dimensional random vector $Y^n$ over the alphabet $\Ycal^n$. The distribution of $Y^n$ depends on an underlying signal $\theta \in \Theta$ through a conditional distribution kernel $P_{Y^n|{\theta}}$. Here $\theta$ represents a vector of arbitrary dimension, where this dimension can be fixed or otherwise change with $n$. We assume that for almost every $\theta$, $\{Y_n\}_{n\in \mathbb N}$ is a stationary ergodic process. 
%
The encoder compresses the data $Y^n$ using $nR_n$ bits. The decoder produces a compressed representation
$Z^n$ of $Y^n$ over the alphabet $\Zcal^n$, which is embedded in the metric space $(\Zcal^n,\nu)$. We assume that the metric spaces $(\Theta,\rho)$ and $(\Zcal^n,\nu)$ are complete and separable, i.e., Polish. \par
For an estimator $\phi : \Zcal^n \to \Theta$ measurable with respect to the Borel $\sigma$-algebras on $(\Zcal^n,\nu)$ and $(\Theta,\rho)$, define its $p$-th CE risk as
 \begin{equation}
  D^{\CE}_\phi (\theta) 
  \equiv 
  \label{eq:CE_risk}
  \ex{ \rho^p(\theta, \phi(Z^n))}, \qquad p\geq 1, 
\end{equation}
where the expectation is with respect to $P_{Y^n|\theta}$ and possible randomness in the encoding and decoding $Y^n \to Z^n$ as explained below. Whenever a prior $P_{\theta}$ on $\Theta$ is given, the expectation above is also with respect to this prior. Compared to our earlier discussion, the loss function $\ell$ in $\eqref{eq:true_risk}$ is the $p$-th power of the metric $\rho$ in the notation of \eqref{eq:CE_risk}. Our setting is restricted to loss functions of this form. \par
The goal of this paper is to characterize the CE risk
\eqref{eq:CE_risk}. Our approach to this characterization requires additional definitions and assumptions to follow. 

\subsection{Compression Setup}
For integers $n$ and $M_n$, a \emph{lossy-compression $n$-block code} $(f_{\enc},f_{\dec},M_n)$, or simply a $n$-block code, is defined as the pair of mappings
\eas{
 & f_{\enc} : \Ycal^n \to \{1,\ldots,M_n\} \\
 & f_{\dec} : \{1,\ldots,M_n\} \to \Zcal^n.
}
We denote by $R_n \equiv \log(M_n)/n$ the rate of the code in bps.
In the following, we consider codes for which $f_{\enc}$ and $f_{\dec}$ are random such that the mapping
\begin{equation}
    \label{eq:code_def}
Y^n \to Z^n = f_{\dec}(f_{\enc}(Y^n))
\end{equation}
is measurable\footnote{The $\sigma$-algebra on $\Ycal^n$ is implicit in our notation. In contrast, for $(\Zcal^n,\nu)$ and $(\Theta,\rho)$ we use their Borel $\sigma$-algebras.}. In this case, it is convenient to refer to the code $(f_{\enc},f_{\dec},M_n)$ by the transition probability kernel $P_{Z^n|Y^n}$ it induces on $\Yc^n \times \Zc^n$. Namely, for a Borel measurable $A \subset \Zcal^n$,
\begin{align*}
& P_{Z^n|Y^n}(Z^n\in A|Y^n=y^n)  \\ & \qquad \equiv 
\Pr\left[f_{\dec}(f_{\enc}(Y^n)) \in A | Y^n = y^n\right]. 
\end{align*}
The \emph{output distribution} of the code $P_{Z^n|Y^n}$ is its marginal distribution $P_{Z^n}$ over $\Zcal^n$, i.e.,
\[
P_{Z^n}(\diff z^n) \equiv \int_{\Ycal^n} P_{Z^n|Y^n}(\diff z^n|y^n)P_{Y^n}(\diff y^n). 
\]

In the next section, we provide standard notations from rate-distortion theory required for defining the RD risk, a would-be asymptotic characterization of $D_{\phi}^{\CE}$. 

\subsection{Rate-Distortion Setup}
The distortion between $Y^n$ and $Z^n$ is measured using the function 
$d : \Ycal^n \times \Zcal^n \to [0,\infty)$, which is further assumed sub-additive in the sense that 
\[
d(y^{k+m},z^{k+m}) \leq d(y^k,z^k) + d(y_{k+1}^{m+k},z_{k+1}^{m+k}),
\]
for any $y^{m+k} \in \Ycal^{m+k}$, $z^{m+k} \in \Zcal^{m+k}$ with $k\leq m$, where $u_k^m = (u_k,\ldots,u_m)$. We also set \[
d_{\min} \equiv \inf_{y\in\Ycal,z\in\Zcal} d(y,z),
\]
and note that $\inf_{z^n,y^n} d(y^n,z^n) \leq n \cdot d_{\min}$ by sub-additivity. \par
For a target distortion $D_y \geq d_{\min}$ and a distribution $P_{Y^n}$ on $\Yc^n$, the RD function is defined as \cite{berger1971rate} 
\begin{align}
    \label{eq:RDF_def}
\Rcal_{Y^n}(D_y) \equiv \min_{\substack{P_{Z^n|Y^n} \\
\ex{d(Y^n,Z^n)} \leq D_y n }} I(Y^n ; Z^n),
\end{align}
where $I(Y^n ; Z^n)$ is the mutual information of $P_{Y^nZ^n}$ and the minimum is over all conditional distribution kernels $P_{Z^n|Y^n}$ satisfying the prescribed distortion constraint. 
We also define the single-letter RD function with respect to $Y^n$ as 
\ea{
R(D_y) = \inf_n \frac{1}{n} \Rcal_{Y^n}(D_y) = \lim_{n\to \infty} \frac{1}{n}\Rcal_{Y^n}(D_y).
\label{eq:rate dist fun}
}
The last transition is due to
sub-additivity of $d$ and the stationarity of $Y^n$ \cite[Lem. 10.6.2]{gray2011entropy}. \par
The single-letter DR function $D_y(R)$ is the inverse function of $R(D_y)$ in \eqref{eq:rate dist fun} when $R$ is smaller than the entropy rate of $Y^n$ and zero otherwise. 
\par
A code $(f_\enc,f_\dec,M)$ is said to be $D_y$-\emph{admissible} if its expected distortion satisfies
\[
\ex{d\left(Y^n,Z^n\right)} \leq n D_y. 
\]
The fundamental result of source coding theory says that, for a wide class of distributions $P_{Y^n}$, 
there exists a sequences of codes $P_{Z^n|Y^n}$ such that
\begin{align}
\begin{split}
    \label{eq:good_codes_def}
     \ex{d\left(Y^n,Z^n \right)} \leq n D_y \quad \text{while} \quad
     \lim_{n\to \infty} R_n = R(D_y). 
    \end{split}
\end{align}
A sequence of codes satisfying \eqref{eq:good_codes_def} is referred to as a \emph{good} rate-distortion code for $(Y^n,d)$ at distortion level $D_y$ \cite{kanlis1997compression, shamai1997empirical, weissman2005empirical, schieler2013connection, KostinaVerdu}. \par
Throughout this paper we assume the following:
\begin{assumption}{A-RD}{\textnormal{(uniqueness of RD-achieving distribution).}}
\label{ass:A-RD}
The minimum in \eqref{eq:RDF_def} is achieved by a unique $P_{Z^{* n}|Y^n}$, which also satisfies $\ex{d(Y^n,Z^n)}=nD_y$. 
\end{assumption}
The notation \ref{ass:A-RD} indicates that Assumption~\ref{ass:A-RD} only concerns properties of the RD setup. 
Excluding pathological cases of limited interest, Assumption \ref{ass:A-RD} holds whenever $0\leq d_{\min} \leq D_y \leq \inf_{z^n} \ex{d(Y^n,z^n)}$, i.e., for distortion levels $D_y$ at which the source coding problem with respect to $Y^n$ is non-trivial. Under Assumption~\ref{ass:A-RD}, Kostina and Verd\'u \cite{KostinaVerdu}
provided a characterization of the conditional relative entropy between any $D_y$-admissible code $P_{Z^n Y^n}$ and the RD-achieving distribution $P_{Z^{* n} Y^n}$.
%
%
\begin{prop}[Properties of good RD codes {\cite[Thm. 1]{KostinaVerdu}}]
	\label{prop:KostinaVerdu}
	Let $P_{Z^n|Y^n}$ be a code such that $\ex{d(Y^n,Z^n)}\leq nD_y$ for some $d_{\min} \leq D_y$. Suppose that the support of $P_{Z^n}$ is a subset of the support of $P_{Z^{* n}}$ and that 
	Assumption~\ref{ass:A-RD} holds. Then
	\[
	D\left( P_{Y^n Z^n}  \parallel  P_{Y^n|Z^{*n}} P_{Z^n} \right) = I(Y^n; Z^n) - \Rcal_{Y^n}(D_y),
	\]
	for $\Rcal_{Y^n}(D_y)$ in \eqref{eq:RDF_def}.
\end{prop}
We recall that the relative entropy between two distributions $P_U$ and $P_V$ on $\Xcal$, with $P_U \ll P_V$, is defined as 
\[
D(P_U\parallel P_V) \equiv \int_{\Xcal} \log \left(\frac{\diff P_U}{\diff P_V}(x) \right) \diff P_U(x),
\]
where $\frac{\diff P_U}{\diff P_V}$ is the Radon-Nikodym derivative of $P_U$ with respect to $P_V$. 
\par

In addition to Assumption~\ref{ass:A-RD}, our main results utilize the following two assumptions for a sequence of codes $\{P_{Z^n|Y^n} \}_{n=1}^\infty$. We discuss each assumption in more detail in the specific setting where it is used. 
\begin{assumption}{A-C1}{\textnormal{ (conditional absolute continuity).}}
\label{ass:A-C1}
For almost every $y^n \in \Ycal^n$, $P_{Z^{n}|Y^n=y^n}$ is absolutely continuous with respect to $P_{Z^{*n}|Y^n=y^n}$.
\end{assumption}
\begin{assumption}{A-C2}{\textnormal{(convergence of output distributions).}}
\label{ass:A-C2} The sequence of code output distributions $\{P_{Z^n}\}_{n=1}^\infty$ and the sequence of RD-achieving distribution marginals $\{P_{Z^{*n}}\}_{n=1}^\infty$ satisfy
\[
\lim_{n\to \infty} \frac{1}{n}D(P_{Z^n}\parallel P_{Z^{* n}}) = 0.
\]
\end{assumption}
The notations \ref{ass:A-C1} and \ref{ass:A-C2} indicate that these assumptions concern properties of the compression code. 

%
\subsection{The Rate-Distortion Risk}
For an estimator $\phi : \Zcal^n \to \Theta$ measurable with respect to the Borel $\sigma$-algebras on $(\Zcal^n,\nu)$
and $(\Theta,\rho)$, define its $p$-th RD risk as
\begin{equation}
D_{\phi}^*(R,\theta) \equiv D^*(R,\theta,\phi,d,\rho^p) \equiv  \ex{\rho^p\left(\theta,\phi(Z^{*n})\right)},
\label{eq:RD_risk}
\end{equation}
where the expectation is with respect to $P_{Z^{*n},Y^n|\theta}$. Note that $D^*_\phi(R,\theta)$ depends on $R$ since it depends on $P_{Z^{*n}Y^n}$, the minimizer of \eqref{eq:RDF_def}. Whenever a prior $P_\theta$ is provided, we consider the Bayes RD risk of $\phi$ 
\begin{align}
\label{eq:BayesRD_def}
D_{\phi}^*(R) \equiv \mathbb E_{\theta} \left[D^*_\phi(R,\theta) \right], 
\end{align}
(the expectation in \eqref{eq:BayesRD_def} is with respect to $P_\theta P_{Y^n|\theta}P_{Z^{*n}|Y^n}$) and the (optimal) Bayes RD risk
\begin{equation}
    \label{eq:Bayes_optimal_RD_risk}
D^*(R) \equiv \inf_{\phi} D_{\phi}^*(R).
\end{equation} 
Our main result in Section~\ref{sec:main} establishes a general connection between $D_{\phi}^*(R,\theta)$ or $D_{\phi}^*(R)$ and the CE risk of \eqref{eq:CE_risk}. Later in this paper, in Sections~\ref{sec:discrete} and \ref{sec:Gaussian}, we focus on two special settings of the problem formulation above. These are described next, along with specific examples. 

\subsection{Special Case I: Discrete Memoryless Setting \label{subsec:discrete}}
In this setting, we assume that $\theta = \theta^n=(\theta_1,\ldots,\theta_n)$ represents $n$ independent samples from a distribution $P_{\theta_1}$ with support $\Theta_1$, $\Theta = \Theta_1^n$, and $P_{Y^n|\theta^n}$ factorizes as 
\[
P_{Y^n|\theta^n} (y^n,\theta^n)= \prod_{i=1}^n P_{Y_1|\theta_1} (y_i,\theta_i). 
\]
Additional assumptions for this section are as follows:
\begin{itemize}
    \item[~~(i)] $\Zcal$ is discrete.
    \item[~(ii)] Single-letter bounded loss function: $\rho(\theta^n,\hat{\theta}^n)=  \frac{1}{n} \sum_{i=1}^n \rho(\theta_i,\hat{\theta}_i)$, where $\sup_{\theta_1,\hat{\theta}_1}\rho(\theta_1,\hat{\theta}_1)\leq M$, for a universal constant $M>0$. 
\item[(iii)] Single-letter distortion function: $d(y^n,z^n) = \sum_{i=1}^n d(y_i,z_i)$.
\end{itemize}
Our setting of Figure~\ref{fig:system} under these assumptions is denoted as \emph{discrete memoryless setting}; we analyze this setting 
in Section~\ref{sec:discrete}. Since $\Zcal$ is discrete, the normalized Hamming distance 
\ea{
\rho_H(u^n,v^n) \equiv  \frac{1}{n} \sum_{i=1}^n \mathbf 1_{\{u_i \neq v_i\}},
\label{eq:Hamming}
}
is a natural metric on this space. When $\nu = \rho_H$ and $p=1$, a TC inequality suitable to our need was given by Marton \cite{marton1986simple,marton1996bounding}; see Proposition~\ref{prop:Marton} below. Our main results for this setting (see Theorem~\ref{thm:main_product}) uses this TC inequality to control the difference between the CE risk and the RD risk of estimators $\phi : (\Zcal^n, \rho_H) \to (\Theta_1^n,\rho)$.
\par
A simple example that we explore in this setting assumes that $\theta^n$ is an iid binary sequence, the data $Y^n$ is obtained by passing $\theta^n$ through a binary symmetric channel, and all metrics and the distortion function are the normalized Hamming distance. Namely, we have $\Theta = \{0,1\}^n$, $\Zcal = \Ycal = \{0,1\}$, and $\rho = \nu = d = \rho_H$.

\subsection{Special Case II: Quadratic Gaussian Setting}

In Section~\ref{sec:Gaussian}, we focus on the case where the data has a multivariate normal distribution 
\[
Y^n \sim \Ncal(\mu^n, \Sigma_n),\qquad \mu^n \in \reals^n, \quad \Sigma_n \in \mathbf S_{+}^n.
\]
We also assume that $d$ and $\nu$ are the normalized Euclidean norm. For this setting, we use the TC inequality of Talagrand \cite{talagrand1996transportation}; see Proposition~\ref{prop:Talagrand} below. In principle, attaining the RD function of $Y^n$ requires knowledge of $\mu^n$ and $\Sigma_n$. These parameters may be provided as side information or estimated at the encoder with no additional performance cost if their entropy rate in $n$ is zero. 
\par
A specific example we consider under this setting is the multiple Gaussian observation setting, a generalization of the bivariate Gaussian observation setting \eqref{eq:2d_Gaussian_model} to an arbitrary number of observations.


\section{Transportation-Theoretic Interlude and Preliminary Results \label{sec:TC}}
In this section, we review transportation theoretic notions and derive a preliminary characterization of the difference between the CE and RD risks.  

\subsection{Wasserstein Distance}
Transportation-cost information inequalities are expressed through the Wasserstein distance. We now review the definition of this distance and related relevant results. We refer to \cite{djellout2004transportation, raginsky2013concentration,villani2008optimal} for detailed background and applications of transportation theory.  
%

\begin{definition}[Wasserstein distance]
	\label{def:Wasserstein distance}
	Let $P_U$, $P_{U'}$ be two Borel probability measures with respect to a metric $\nu$ on the Polish space $\Uc$. The $p$-Wasserstein distance between $P_U$ and $P_{U'}$, for $p\geq 1$, is defined as
	\[
	W_p(P_U,P_{U'}) \equiv \left(\inf_{P_{UU'}} \ex{\nu^p(U,U')} \right)^{1/p},
	\]
	where the infimum is over all joint probability distributions $P_{UU'}$ on the product space $\Ucal \times \Ucal$ with marginals $P_U$ and $P_{U'}$. 
\end{definition}
We also define the $p$-Wasserstein distance between $P_U$ and $P_{U'}$ conditioned on a third probability measure $P_V$ by
\begin{align}
W_p(P_{U|V},P_{U'|V}|P_V) \equiv \left(\mathbb E\left[{W_p^p(P_{U|V},P_{U'|V})}\right]\right)^{1/p},
	\label{eq:cond wass}
\end{align}
where $P_{UV}$ and $P_{U'V}$ are over $\Uc\times \Vc$ and the expectation is with respect to $P_V$. Note the identity
\[
W_p(P_{U|V},P_{U'|V}|P_V) = \left(\inf_{P_{UU'V}} \ex{\nu^p(U,U')} \right)^{1/p},
\]
where the infimum is over all joint probability distributions $P_{UU'V}$ over $\Ucal\times \Ucal \times \Vcal$ with marginals $P_{UV}$ and $P_{U'V}$. 
\par 
The following continuity argument provides the main motivation for using the Wasserstein distance in our setting. 
\begin{prop}[continuity of the risk in $W_p$]\label{prop:Lipschitz}
Let $P_{XY}$ be a distribution on $\Xc \times \Yc$ and let $P_{Z|Y}$ and $P_{Z'|Y}$ be two conditional distribution kernels from $\Yc$ to $\Zc$. For any $L$-Lipschitz $\phi : (\Zc,\nu) \to (\Xc,\rho)$ and $p \geq 1$ such that $
\ex{ \rho^p \left(X , \phi(Z) \right) } < \infty$ and $\ex{ \rho^p\left(X , \phi( Z' )\right) } < \infty$, we have
\begin{align}
    & \left| \left(\ex{ \rho^p \left(X , \phi(Z) \right) }\right)^{1/p} -
    \left(\ex{ \rho^p\left(X , \phi( Z' )\right) } \right)^{1/p}
    \right| \nonumber \\
    & \qquad \qquad \leq 
     L \cdot W_p(P_{Z|Y}, P_{Z'|Y} | P_Y). \label{eq:Wass_cont}
\end{align}
\end{prop}
\begin{proof}
For any Markov chain $X \to Y \to (Z,Z')$, the triangle inequality and Lipschitz continuity of $\phi$ imply
\ea{
& \left( \ex { \rho^p \left(X , \phi(Z)\right)} \right)^{1/p} \leq 
    \left(\ex{ \rho^p \left(X, \phi( Z')\right) } \right)^{1/p} \label{eq:Wass_cont_proof} \\
    & \qquad +  \left( \ex{ \rho^p \left( \phi(Z'), \phi(Z) \right) } \right)^{1/p} \nonumber \\
    & \leq 
    \left(\ex{ \rho^p \left(X , \phi(Z')\right)} \right)^{1/p}+ L \left(\ex{ \nu^p ( Z', Z)) } \right)^{1/p}. \nonumber
}
One side of the inequality in \eqref{eq:Wass_cont} is now obtained by taking the minimum over all joint distributions $P_{YZZ'}$ with marginals $P_{YZ}$ and $P_{YZ'}$. 
The second side of the inequality in \eqref{eq:Wass_cont} follows by interchanging the role of $Z$ and $Z'$ in \eqref{eq:Wass_cont_proof}. 
\end{proof}

\subsection{Transportation-Cost Inequality}
A probability measure on a metric space whose Wasserstein distance with respect to any other measure is bounded by the relative entropy is said to satisfy a transportation cost (TC) inequality.
\begin{definition}[transportation-cost inequality {\cite[Def. 3.4.2]{raginsky2013concentration}}]
Let $(\Xcal,\rho)$ be Polish and fix $p \geq 1$. 
It is said that a probability measure $Q$ on $(\Xcal,\rho)$ satisfies the TC inequality with constant $c>0$ if
\ea{
W_p(P,Q) \leq \sqrt{ 2c D(P\parallel Q)},
\label{eq:TC def}
}
for every Borel probability measure $P$ on $(\Xcal,\rho)$. 
When \eqref{eq:TC def} holds,  we say that $(\Xcal,\rho,Q)$ is $\TC(c)$. 
\end{definition}
In \cite{1057176, marton1996bounding}, Marton showed that any product measure on a discrete space is $\TC(1/4n)$ with respect to the normalized Hamming distance of this space. 
\begin{prop}[\cite{marton1996bounding}]
	\label{prop:Marton}
	Consider the Polish space  $(\Zcal^n,\rho_H)$ where $\Zcal$ is a countable set and 
	$\rho_H$ is the normalized Hamming distance. For any probability measure $P_{Z^n}$ on $\Zcal^n$, 
	\begin{align}
	\label{eq:trans_cost_product}
	W_1(Q_{Z^n},P_{Z^n}) \leq \sqrt{\frac{1}{2n} D(P_{Z^n}\parallel Q_{Z^n}}). 
	\end{align}
\end{prop}

With some abuse of notation, denote by $\Ncal(\mu^n,\Sigma_n)$ the Gaussian measure on $\reals^n$ centered at $\mu^n$ with $\Sigma_n \in \PSD$. In \cite{\cite{talagrand1996transportation}}, Talagrand showed that the standard Gaussian measure is $\TC(1/n)$ with respect to the normalized Euclidean distance in $\reals^n$. 
\begin{prop}[\cite{talagrand1996transportation}] \label{prop:Talagrand}
     Consider the metric space $\left(\reals^n,\frac{1}{\sqrt{n}}\|\cdot\|_2\right)$, where $\|\cdot\|_2$ is the Euclidean norm. Suppose that the measure $P$ on $\reals^n$ has a density. Then 
	\[
	W_2\left(P, \Ncal({\bf 0},\Id_n)\right) \leq \sqrt{\frac{2}{n} D\left( P \parallel \Ncal({\bf 0},\Id_n) \right)}.
	\]
\end{prop}
In Section~\ref{sec:Gaussian} below we use the following extension of Proposition~\ref{prop:Talagrand} to general real Gaussian measures. The proof is in Appendix~\ref{app:Talagrand}.
\begin{prop} \label{prop:Talagrand_ext} 
	Consider the space $\left(\reals^n,\frac{1}{\sqrt{n}}\|\cdot\|_2 \right)$, where $\|\cdot\|_2$ is the Euclidean distance. Suppose that the measure $P$ on $\reals^n$ has a density. Then 
	\begin{equation}
	W_2\left(P, \Ncal(\mu^n,\Sigma_n)\right) \leq \sqrt{\frac{2}{n}\left\|\Sigma_n\right\| D\left( P \parallel \Ncal(\mu^n,\Sigma_n) \right)}.
	\label{eq:Talagrand_ext}
	\end{equation}
\end{prop}

\subsection{Transportation-cost and the RD risk \label{sec:main}}
%
Under the setting of Figure~\ref{fig:system}, consider a distribution $P_{\theta Y^n}$ on $\Theta \times \Yc^n$, a sub-additive distortion measure $d$ on $\Yc^n \times \Zc^n$, and a distortion level $D_y \geq d_{\min}$ such that Assumption~\ref{ass:A-RD} holds. 
The following lemma provides a 
general bound on the difference between the CE risk \eqref{eq:CE_risk} and the RD risk \eqref{eq:RD_risk}. The proof is in Appendix~\ref{app:proof:main}.
\begin{lem} \label{lem:main}
Fix $p\geq 1$. Suppose that for $P_{Y^n}$-almost every $y^n\in \Ycal^n$, there exists $c=c(y^n)>0$ such that $P_{Z^{*n}|Y^n=y^n}$ is $\TC(c)$.
Let $P_{Z^n|Y^n} : \Yc^n \to \Zc^n$ be a $D_y$-admissible random code satisfying Assumptions \ref{ass:A-RD} and \ref{ass:A-C1}. For $\phi: (\Zc^n,\nu) \to (\Theta,\rho)$ that is $L$-Lipschitz, its CE risk $D_\phi^{\CE}$ of \eqref{eq:CE_risk}, and its RD risk $D^*_{\phi}$ of \eqref{eq:RD_risk}, satisfy
\begin{align}
\label{eq:main_2}
     & \left| 
     \left( D^{\CE}_\phi (\theta) \right)^{1/p}
     -\left( D_{\phi}^*(R(D_y),\theta) \right)^{1/p}
     \right| \\
     & \leq 
     L \sqrt{2 n \ex{c(Y^n)} \left(R_n - R(D_y) + \frac{1}{n} D(P_{Z^n}\parallel P_{Z^{*n}})\right)}, \nonumber 
\end{align}
provided $\ex{\rho^p\left(\theta,\phi(Z^n) \right)}<\infty$ and $D^*_{\phi}(R,\theta) < \infty$.
\end{lem}
Lemma~\ref{lem:main} bounds the difference between the risk of an estimator of $\theta$ from the compressed data and the RD risk of this estimator in terms of its Lipschitz constant the expected TC constant of the measure $P_{Z^{* n}|Y^n=y^n}$ over $(\Zcal^n, \nu)$. \par
The bound in \eqref{eq:main_2} is general as it depends on the task of estimating $\theta$ only through the Lipschitz constant of the estimator. Next, in Sections~\ref{sec:discrete} and \ref{sec:Gaussian}, we explore special cases where $L = \Ocal(1)$ and $\ex{c(Y^n)} = \Ocal(1/n)$. For such cases, Lemma~\ref{lem:main} provides sufficient conditions for the convergence
\begin{align*}
     \left| 
     \left(
     D_\phi^{\CE}(\theta)
     \right)^{1/p}  -
    \left(D_{\phi}^*(R(D_y),\theta) \right)^{1/p}
    \right| \to 0,\quad {\text{as $n\to \infty$}.}
\end{align*} 

\section{Discrete Memoryless Setting \label{sec:discrete}}
In this section, we focus on the discrete memoryless setting presented in Section~\ref{subsec:discrete}.
It is well known that the RD function of $Y^n$ in this case is \cite{ThomasCover}
\begin{equation}
    \label{eq:RDF_sep}
R(D_y) = \inf_{\substack{P_{Z_1|Y_1} \\
\ex{d(Y_1,Z_1)} \leq D_y}} I(Y_1 ; Z_1),
\end{equation}
and $P_{Z^{*n}Y^n}$ satisfying \eqref{eq:RDF_def} with equality decomposes as
\[
P_{Z^{* n}Y^n} (z^n,y^n)= \prod_{i=1}^n P_{Z^*_i Y_i}(z_i,y_i).
\]
Consequently, the Bayes estimator of $\theta = \theta^n$ from $Z^{*n}$ with respect to $\rho$ is of the form $\phi^B(z^n) = \left(\phi^{B}_1(z_1),\ldots,\phi^{B}_n(z_n) \right)$, where
\begin{equation}
    \label{eq:Bayes_estimator}
\phi^{B}_i(z) = 
\argmin_{t \in \Theta_1}~\ex{ \rho\left(\theta_i,t \right) \mid Z^*_i = z }.
\end{equation}
Combining these facts with Proposition~\ref{prop:Marton} and Lemma~\ref{lem:main}, we conclude the following result. The proof is provided in Appendix~\ref{proof:thm:main_product}. 
\begin{thm}
\label{thm:main_product}
Fix $d_{\min} \leq D_y$ such that Assumption \ref{ass:A-RD} holds. Suppose that $P_{Z^n|Y^n} : \Ycal^n \to \Zcal^n$ is a $D_y$-admissible code of bitrate $R_n$ and that $R_n \to R$. 
Assume that $\{P_{Z^n|Y^n} \}_{n=1}^\infty$ satisfies Assumptions \ref{ass:A-C1} and \ref{ass:A-C2}. 
 \begin{itemize}
     \item [(i)] For any estimator $\phi : \Zcal \to \Theta_1$, 
\begin{align}
    \label{eq:main_prod_1}
 \lim_{n\to \infty} \frac{1}{n} \sum_{i=1}^n \ex{\rho\left(\theta_i,\phi\left(Z_i\right) \right)} & =  \ex{\rho\left(\theta_1,\phi(Z^*_1) \right)} \nonumber \\
 & = 
 D^*_{\phi}(R(D_y)). 
\end{align}
In particular, for the Bayes estimator $\phi^B$ of \eqref{eq:Bayes_estimator},
\begin{align*}
  \lim_{n \to \infty} \mathbb E \left[ D_{\phi^B}^{\CE}(\theta^n) \right] & = D^*(R(D_y)),\quad \text{as $n\to \infty$}.
 \end{align*}
\item[(ii)] For a sequence of Lipschitz estimators $\phi_n : (\Zcal^n,\rho_H) \to (\Theta,\rho)$ whose Lipschitz constants are uniformly bounded in $n$, set 
$\hat{\theta}^n =\phi_n\left(Z^n \right)$. Then
\begin{align}
\label{eq:main_prod_2}
\liminf_{n\to \infty} \frac{1}{n} \sum_{i=1}^n \ex{\rho\left(\theta_i, \hat{\theta}_i \right)} \geq  
D^*(R(D_y)). 
\end{align}
 \end{itemize}
\end{thm}
%
\subsection{Discussion}
Theorem~\ref{thm:main_product} provides both achievability and converse coding result under the CE scenario. Specifically, (i) in Theorem~\ref{thm:main_product} says that the CE risk of any scalar estimator converges to its RD risk, hence the Bayes RD risk 
$D^*(R)$ is achievable by employing the Bayes estimator \eqref{eq:Bayes_estimator} at each coordinate. (ii) in Theorem~\ref{thm:main_product} says that any Lipschitz estimator with respect to the normalized Hamming distance on $\Zcal^n$ cannot attain asymptotic risk smaller than the Bayes RD risk. In particular, among all Lipschitz estimators of $\theta^n$ from the compressed representation $Z^n$, asymptotically,
the Bayes estimator $\phi^B$ of \eqref{eq:Bayes_estimator} has the minimal CE risk. \par
These results relies on 
Assumptions \ref{ass:A-C1} and \ref{ass:A-C2}. Under the discrete memoryless setting, Assumption \ref{ass:A-C1} is relatively mild since a code that satisfies it can be obtained from any good RD code with little modifications \cite{kanlis1997compression}. On the other hand, \ref{ass:A-C2} is much more restrictive; examples for good RD codes violating it are provided in \cite{KostinaVerdu} and in \cite[Prop. 2]{kanlis1996typicality}. These examples suggest that it is challenging to construct a good RD code that satisfies \ref{ass:A-C2} without randomness in its construction. Nevertheless, \ref{ass:A-C2} is known to hold in the following cases:
\begin{enumerate}
	\item[(i)] The codewords are randomly drawn from $P_{Z^{*n}}$; encoding is done using joint typicality \cite{ThomasCover}.
	\item[(ii)] The output distribution of the code $P_{Z^n}$ is uniform over the codewords \cite{schieler2013connection}.
\end{enumerate}
\par

Next, in Section~\ref{subsec:binary}, we evaluate in closed-form the RD risk under a particular discrete memoryless model. We also compare this risk to the optimal source coding performance, described by the indirect RD function. Following the interpretation of the RD risk provided by Theorem~\ref{thm:main_product}, this comparison characterizes the price of using the CE scheme instead of the optimal compression and estimation scheme.

\subsection{Example: Binary Signal under Bit-flip Noise
\label{subsec:binary}}
Let $\Bcal(\pi)$ indicate the Bernoulli distribution with probability of success $\pi$. 
Consider the case where $\theta = \theta^n \simiid \Bcal(\pi)$ and
\ea{
Y_i = \theta_i \oplus W_i, \quad i=1,\ldots,n,
\label{eq:binary_obsv_model}
}
where $W^n \simiid \Bcal(\alpha)$ is independent of $\theta^n$ and $\oplus$ denotes addition modulo 2. For simplicity, we assume that $\alpha, \pi \leq 1/2$, and note that the general case is obtained by replacing $\alpha$ with $\min\{\alpha, 1-\alpha\}$ and $\pi$ with $\min\{\pi, 1-\pi\}$ throughout. We have that $Y^n \simiid \Bcal(\pi \star \al)$, where $\star$ indicates the convolution of binary measures:
\[
\pi \star \al \equiv \pi(1-\alpha)+\alpha(1-\pi). 
\]
We also take $\rho(\theta,\hat{\theta})$ and $d(y,z)$ to be the normalized Hamming distance $\rho_H$ of \eqref{eq:Hamming}. 
To evaluate the Bayes RD risk, note that the RD function of $Y^n$ at a distortion level $0 < D_y < \pi \star \al$ is given by
\ea{
R(D_y) = 
\begin{cases}
  h(\pi \star \alpha) - h(D_y) & 0 <   D_y < \pi \star \alpha \\
 0 & \pi \star \alpha \leq D_y,
\end{cases}
\label{eq:R D binary}
}
where $h(x)$ is the binary entropy function. The conditional distribution $P_{Z^*_1|Y_1}$ attaining $R(D_y)$ in \eqref{eq:RDF_sep} is given by the channel \cite{ThomasCover}
\begin{equation}
\label{eq:binary_back_channel}
Y_1 = Z_1^* \oplus V_1,
\end{equation}
where $V_1 \simiid \Bcal(D_y)$ is independent of $Z^*_1$. 
The Bayes and ML estimators of $\theta_1$ from $Z^*_1$ coincides and equal $\phi(z) = z$. The risk of $\phi$ equals $\alpha \star D_y$, the probability of a bit-flip in the channel from $\theta_1$ to $Z^*_1$. 
Denote by $h^{-1}$ the inverse of $h(x)$ in the range $[0,1/2]$. Using \eqref{eq:R D binary} to write $D_y$ as a function of $R$, we get
\begin{align}
D^*(R) & = \alpha \star D_y = \alpha \star h^{-1}\left(\left[h(\pi \star \alpha)-R\right]^+\right) \nonumber \\
& =
\alpha +  (1-2\alpha) h^{-1}([h(\pi \star \alpha)-R]^+).
 \label{eq:ce_binary}
\end{align}
Figure~\ref{fig:iRDF_iid_bound} depicts the function $D^*(R)$ in \eqref{eq:ce_binary}.
\par
For comparison, the indirect DR function $D_{\ind}(R)$ in the setting of this example was considered in \cite{KipnisRini2015}, which provided a parametric expression for evaluating it. This parametric expression has a closed form only in the case $\pi=1/2$, as
\begin{align*}
D_{\ind}(R) = \alpha + 
    (1-2\alpha) h^{-1}\left(\left[1 - R\right]^+ \right),\quad \pi=1/2.
\end{align*}
Since $h(1/2 \star \alpha) = 1$, the indirect DR function and the RD risk coincide for $\pi=1/2$. On the other hand, \cite{KipnisRini2015} implies that the indirect DR function is strictly smaller than the RD risk for $ 0 < \alpha < \pi < 1/2$.  Figure~\ref{fig:iRDF_iid_bound} illustrates the relation between the two risks for a specific choice of $\alpha$ and $\pi$ in the range 
$ 0 < \alpha < \pi < 1/2$. 
For this choice, $\alpha$ is the minimal Hamming loss, no matter the available compression bitrate. In particular, both $D^*(R)$ and $D_{\ind}(R)$ attains this minimal loss for $R \geq h(\pi \star \alpha)$. Indeed, $h(\pi \star \alpha)$ is the entropy-rate of $Y^n$, and both the CE and EC scenarios essentially compress $Y^n$ almost losslessly at this rate. Finally, we note that while the indirect DR function attains the maximum distortion $D=\pi$ at rate zero, we have that $D^*(R) = \pi$ already at $R_{\gamma} \equiv h(\pi \star \alpha) - h((\pi-\alpha)/(1-2\alpha))$.
This difference is rather interesting as it highlights that the encoded representation of $Y^n$ provides no information on $\theta^n$ for a rate less than $R_{\gamma}$. 

\begin{figure}
\begin{center}

\begin{tikzpicture}
    \begin{axis}[
    width=8cm,
    height=6cm,
    legend style={at={(1,1)},
      anchor=north east, legend columns=1},
    ylabel={$\rho_H$},
    xlabel={$R$ [bps]},
    ytick={0,0.05,0.25},
    yticklabels={0,\scriptsize $\alpha$,.25},
    xtick={0,0.08,0.25,0.5,0.75,0.81,0.845,1},
    xticklabels={0,,.25,.5,,\scriptsize{$h(\pi)$},,1},
    legend cell align={left},
    ymin=0,
    xmin=0,
    xmax=1,
    ymax=.3,
    ]

\def\Pie{0.25}
\def\Al{0.05}

\addplot[domain=\Al:\Pie, color=red, samples=100, style=ultra thick] 
({0.848-(-((x-.05)/(1-2*0.05))*ln(((x-.05)/(1-2*0.05)))/ln(2) - (1-((x-.05)/(1-2*0.05)))*ln(1-((x-.05)/(1-2*0.05)))/ln(2))},{x});
\addlegendentry{\scriptsize{$D^*(R)$ (Bayes RD risk)}}

\addplot[color = blue, solid, smooth, mark size=1.2pt, domain=0:1, style=ultra thick] table [x=R, y=D, col sep=comma] {./Figs/idrf.csv};

\addlegendentry{\scriptsize{$D_{\ind}(R)$} (optimal) };

\addplot[domain=0.00001:0.25, color=black, samples=100, style=thick, dashed] 
({0.811-(-x*ln(x)/ln(2) - (1-x)*ln(1-x)/ln(2)) },{x});

\addlegendentry{\scriptsize{DR function (optimal w/o noise)}}

\node[] (pi) at (axis cs: 0.2,0.25) {};
\node[] (hpial) at (axis cs: 0.841,0.15) {};

\node[] (minR) at (axis cs: 0.08,0.32) {};

\addplot[dotted, dotted] 
coordinates {(0,.05) (1,.05)};


\addplot[dotted, dotted] 
coordinates {(0.848,0) (0.848, 0.15)};

\addplot[dotted, dotted] 
coordinates {(0,0.25) (0.2, 0.25)};

\addplot[dotted, dotted] 
coordinates {(0.08,0) (0.08, 0.31)};

\end{axis}

\node at (pi) {$\pi$};
\node at (hpial) {\scriptsize $h(\pi \star \alpha)$};
\node at (minR) {\scriptsize $h(\pi \star \alpha) - h\left(\frac{\pi-\alpha}{1-2 \alpha}\right)$};
\end{tikzpicture}



\caption{
Distortion-rate functions for binary data observed under bit-flip noise. The data $Y^n$ is obtained by observing a binary $\pi=0.25$ signal under a binary symmetric channel with bit-flip probability $\alpha = 0.05$. $D^*(R)$ is the Bayes RD risk, attained by compressing $Y^n$ using a good code of bitrate $R$. The optimal distortion-rate trade-off is described by the indirect DR function from \cite{KipnisRini2017ISIT}. The Shannon DR function of a binary $p=0.25$ source, corresponding to the non-noisy case ($\alpha = 0$), is illustrated for comparison.
}
\label{fig:iRDF_iid_bound}
\end{center}
\end{figure}

\section{Quadratic Gaussian Setting \label{sec:Gaussian}}
Suppose that $Y^n \sim \Ncal(\mu^n, \Sigma_n)$ where $\Sigma_n \in \PSD$ and $\mu^n\in \reals^n$ are known parameters. The quadratic ($d = \|\cdot\|_2^2$) DR and RD functions of $Y^n$ are given by the parametric expression \cite{berger1971rate, ThomasCover}:
\eas{
    \label{eq:RDF_Gauss1}
     R(\eta) & \equiv \frac{1}{2n} \sum_{i=1}^n \log^+ \left(\lambda_i(\Sigma_n) /\eta \right) \\
     D_y(\eta) & \equiv \sum_{i=1}^n \min\{\lambda_i(\Sigma_n),\eta\},
     \label{eq:RDF_Gauss2}
}
where $\eta>0$ is the ``waterlevel'' parameter. Furthermore, the distribution $P_{Y^n|Z^{*n}}$ can be represented by the Gaussian channel
\begin{align}
    \label{eq:Gaussian_channel}
    Y^n = Z^{*n} + \Um \Dm W^n,
\end{align}
where $W^n \sim \Ncal\left(0, \Id_n\right)$, $\Um$ is the matrix of right eigenvectors of $\Sigma_n$, and $\Dm = \diag\left[\min\{\lambda_1,\eta\},\ldots,\min\{\lambda_n,\eta\} \right]$. Combining these facts with Lemma~\ref{lem:main}, we obtain the following characterization of the CE risk when $Y^n$ is compressed using a good RD code. The proof is in Appendix~\ref{proof:thm:main_Gaussian1}. 
\begin{thm} \label{thm:main_Gaussian1}
Consider a quadratic Gaussian setting such that $\limsup_n \|\Sigma_n\| < \infty$. Let $\{\phi_n\}_{n=1}^\infty$ be a sequence of estimators $\phi_n : (\mathbb R^n, \frac{1}{\sqrt{n}}\|\cdot\|_2) \to (\Theta, \rho)$ with $\ex{\rho^2(\theta,\phi_n(Z^n))}<\infty$ whose Lipschitz constants are uniformly bounded in $n$. 
Suppose that $P_{Z^n|Y^n} : \reals^n \to \reals^n$ is a $D_y$-admissible code of  bitrate $R_n$, that $R_n\to R(D_y)$, and that $\{P_{Z^n|Y^n}\}_{n=1}^\infty$
satisfies Assumptions \ref{ass:A-C1} and \ref{ass:A-C2}. Then
\begin{align}
    \label{eq:main_Gaussian_2}
 \lim_{n\to \infty}  \left| 
 D_{\phi_n}^{\CE}(\theta)
 -  D_{\phi_n}^{*}(\theta,R(D_y)) \right| = 0.
\end{align}
\end{thm}

\subsection{Discussion}
Theorem~\ref{thm:main_Gaussian1} provides a characterization of the CE risk in terms of the RD risk that is analogous to Theorem~\ref{thm:main_product}. This characterization requires Assumptions \ref{ass:A-C1} and \ref{ass:A-C2} as in the case of Theorem~\ref{thm:main_product}, although here Assumption \ref{ass:A-C1} does not hold when the code is deterministic. Indeed, for such code, $P_{Z^n|Y^n=y^n}$ is a point mass in $\reals^n$ while $P_{Z^{*n}|Y^n=y^n}$ is a Gaussian measure. Nevertheless, we can obtain a code satisfying Assumption \ref{ass:A-C1} from a deterministic code by adding to the compressed representation a small perturbation sampled from a distribution with density in $\reals^n$.
%
As an example for a code that satisfies Assumptions \ref{ass:A-C1} and \ref{ass:A-C2}, consider a codebook consisting of codewords drawn independently from a spherically symmetric distribution with density in $\reals^n$ while encoding is done using maximal cosine similarity \cite{sakrison1968geometric}. 
We note that, by characterizing the performance under random coding, we confirm the existence of deterministic codes satisfying the average performance guarantee over the random ensemble. This is relevant since, in practice, deterministic codes are needed.\par
We now evaluate the RD risks of various estimators in the Gaussian multiple observation setting and compare these risks to the optimal performance described by the indirect DR function.

\subsection{Example: Multiple Gaussian Observation}

Assume that $\rho(\theta, \hat{\theta})=\frac{1}{\sqrt{n}}\left\|\theta - \hat{\theta} \right\|$ and that 
$\theta = \theta^n$ is an $n$-dimensional standard normal vector. Each data point $\Yv_i$ is a vector of $m$ noisy observations of $\theta_i$, as in
\begin{align}
    \label{eq:AWGN}
\Yv_{i,j} = a_j \theta_i + \Wv_{i,j},\quad j=1,\ldots,m, \quad i = 1,\ldots,n,
\end{align}
where $\Wv_{i,j} \simiid \Ncal(0,1)$ and $a_1,\ldots,a_m$ are real numbers. Accordingly, we have $\Ycal = \Zcal = \reals^m$ and $\Theta = \reals^n$. We denote the overall SNR in \eqref{eq:AWGN} by 
\begin{align}
\label{eq:gamma_def}
\gamma \equiv \sum_{j=1}^m a_j^2.
\end{align}
\par
The indirect DR function in the setting of \eqref{eq:AWGN}, describing the minimal distortion under any compression of $Y^n$ to bitrate $R$, is given as \cite[Eq. 10]{gastpar2005lower}:
\begin{equation}
\begin{split} \label{eq:idrf_Gaussian}
D_{\ind}(R) 
& = \frac{1}{1+\gamma} + \frac{\gamma}{1+\gamma} 2^{-2R}.
\end{split}
\end{equation}
\par
We consider the RD risk of two estimators. 
\subsubsection{Maximum Likelihood Estimator}
The ML estimator of $\theta_1$ under \eqref{eq:AWGN} 
\[
\phi^{\ML}(y^m) = \frac{1}{\gamma}\sum_{j=1}^m a_j y_j,
\]
is $1$-Lipschitz and has expected quadratic risk 
\begin{equation}
    \label{eq:risk_ML_standard}
\var\left[\theta_1 - \phi^{\ML}(\Yv_1)\right] = \frac{1}{\gamma}. 
\end{equation}
$\phi^{\ML}$ only depends on the model $P_{\Yv_1|\theta_1}$, hence it may be particularly useful when no assumptions on the compression bitrate $R$ are made.
In view of Theorem~\ref{thm:main_Gaussian1}, the RD risk of $\phi^{\ML}$ describes how \eqref{eq:risk_ML_standard} is affected when the data undergoes lossy compression to bitrate $R$. 
\begin{prop} \label{prop:ML}
Under the setting \eqref{eq:AWGN}, the RD risk of $\phi^{\ML}$ is given by
\begin{align}
\label{eq:ML}
& D^*_{\phi^{\ML}}(R) = \\
& \begin{cases}
        \frac{1}{\gamma} \left(1+2^{-2R}(\gamma-1)\right)
       &  R \leq R_{\gamma} \\
       \begin{multlined}[b]
\left( 2^{-\frac{2}{m}(R-R_{\gamma})-2R_{\gamma}}
         \right)^2 \\
         + \frac{1-2^{-2R}}{\gamma} \left(1-2^{-2R} +2^{-\frac{2}{m}(R-R_{\gamma})}  \right)
\end{multlined}
       & R > R_{\gamma},
     \end{cases}
\end{align}
where 
\[
R_{\gamma} \equiv \frac{1}{2}\log(1+\gamma),
\]
for $\gamma$ as in \eqref{eq:gamma_def}.
\end{prop}
The proof of Proposition~\ref{prop:ML} is in Appendix~\ref{app:proof_Gaussian_ML}. \par
Figure~\ref{fig:Gaussian_centralized} illustrates $D^*_{\phi^{\ML}}(R)$ versus $R$ and $\gamma$ and compare it to $D_{\ind}(R)$ and the Bayes RD risk derived in Proposition~\ref{prop:single_vector} below. It is instructive to consider the asymptotes of $D^*_{\phi^{\ML}}(R)$ in each of these parameters: the case $R\to \infty$ corresponds to removing the compression constraint, in which case $D^*_{\phi^{\ML}}(R)$ converges to \eqref{eq:risk_ML_standard}. The case $\gamma \to \infty$ corresponds to observing $\theta^n$ without noise. In this case,  $D^*_{\phi^{\ML}}(R)$ converge to the standard DR function of $\theta^n$ which is $2^{-2R}$, implying that no performance degradation occurs due to non-optimal compression.
\subsubsection{Bayes Estimator}
The following result provides the Bayes RD risk under \eqref{eq:AWGN}. The proof is in Appendix~\ref{app:proof_gaussian_single}.
\begin{prop} \label{prop:single_vector}
Consider the quadratic Gaussian setting under the model \eqref{eq:AWGN}. The Bayes RD risk is given by
\begin{equation}\label{eq:Gaussian_centralized}
D^{*}(R) = \frac{1}{\gamma+1} + \frac{\gamma }{\gamma + 1} \eta(R),
\end{equation}
where $\eta(R)$ is obtained as
\begin{equation}
\begin{split} \label{eq:single_theta}
\eta(R) = \begin{cases}
  2^{-2R} &  R \leq R_{\gamma} \\
  2^{- \f {2} m (R -R_{\gamma}) - 2R_{\gamma}}
  & R>R_{\gamma}. \\
  \end{cases}
\end{split}
\end{equation}
\end{prop}
\begin{figure*}
\begin{center}
\begin{tikzpicture}
    \begin{axis}[
    width=8cm,
    height=6cm,
    legend style={at={(1,1)},
      anchor=north east, legend columns=1},
    ylabel={MSE},
    xlabel={$R$ [BPS]},
    legend cell align={left},
    ymin=0,
    xmin=0,
    xmax=4,
    ymax=1.05,
    ]

\def\SNR{5}
\def\MM{3}

\addplot[,color=orange, mark = *, mark options={scale=0.2}, style=thick] 
coordinates {(0,2) (1,2)};

\addplot[color=red, style=thick] 
coordinates {(0,2) (1,2)};

\addplot[color=blue, mark=|, mark options={scale=0.5}, style=thick] 
coordinates {(0,2) (1,2)};

\addplot[dashed, color=black, style=thick] coordinates {(0,2) (1,2)};

\addlegendentry{\scriptsize{$D^*_{\phi^{\ML}}$ (ML RD risk)}};
\addlegendentry{\scriptsize{$D^*$ (Bayes RD risk)}};
\addlegendentry{\scriptsize{$D_{\ind}$} (optimal) };
\addlegendentry{\scriptsize{MMSE} (unlimited bitrate) };

\addplot[domain=0:0.5*ln(1+\SNR)/ln(2), color=orange, samples=7, style=thick, mark=*, mark options={scale=0.2}] 
({x},{
(1+2^(-2*x)*(\SNR-1))/\SNR });

\addplot[domain=0.5*ln(1+\SNR)/ln(2):4, color=orange, samples=17, style=thick, mark=*, mark options={scale=0.2}] 
({x},{
2^(-(4/\MM)*(x +(\MM-1)*ln(1+ \SNR)/2/ln(2)))
+((1-2^(-2*x))*(1-2^(-2*x) + 2^(-(2/\MM)*(x-ln(1+ \SNR)/2/ln(2))) ) )/ \SNR 
  });

\addplot[domain=0:4, color=blue, samples=27, style=thick, mark=|, mark options={scale=0.5}] 
({x},{1/(1+\SNR) + \SNR/(1 + \SNR)*2^(-2*x)});

\addplot[domain=0:0.5*ln(1+\SNR)/ln(2), color=red, samples=7, style=thick] 
({x},{1/(1+\SNR) + 2^(-2*x)*\SNR   /(1+\SNR) });

\addplot[domain=0.5*ln(1+\SNR)/ln(2):4, color=red, samples=17, style=thick] 
({x},{
1/(1+\SNR) + \SNR/(1+\SNR) * 2^(-(2*x/\MM + (\MM-1)/\MM * ln(1+\SNR)/ln(2))) });

\addplot[domain=0:4, color=black, samples=27, style=thick, dashed] 
({x},{1/(1+\SNR)});

\node[] (Rcrit) at (axis cs: 1.85,0.5) {};


\addplot[dotted, dotted] 
coordinates {(0.5*ln(1+\SNR)/ln(2),0)  (0.5*ln(1+\SNR)/ln(2),0.48)};

\end{axis}
        
\node at (Rcrit) {\scriptsize $R_0=\frac{1}{2}\log(1+\gamma)$};        
\end{tikzpicture}
\begin{tikzpicture}
    \begin{semilogxaxis}[
    width=8cm,
    height=6cm,
    legend style={at={(1,1)},
      anchor=north east, legend columns=1},
    xlabel={$\gamma~[\mathrm{SNR}]$ },
    legend cell align={left},
    ymin=0,
    xmin=0.5,
    xmax=100,
    ymax=1.05,
    ]

\def\RR{2}
\def\MM{3}

\addplot[,color=orange, mark = *, mark options={scale=0.2}, style=thick] 
coordinates {(0,2) (1,2)};

\addplot[color=red, style=thick] 
coordinates {(0,2) (1,2)};

\addplot[color=blue, mark=|, mark options={scale=0.5}, style=thick] 
coordinates {(0,2) (1,2)};

\addplot[dashed, color=black, style=thick] coordinates {(0,2) (1,2)};

\addlegendentry{\scriptsize{$D^*_{\phi^{\ML}}$ (ML RD risk)}};
\addlegendentry{\scriptsize{$D^*$ (Bayes RD risk)}};
\addlegendentry{\scriptsize{$D_{\ind}$} (optimal) };
\addlegendentry{\scriptsize{MMSE} (unlimited bitrate) };

\addplot[domain=2^(2*\RR)-1:100, color=orange, samples=57, style=thick, mark=*, mark options={scale=0.2}] 
({x},{
(1+2^(-2*\RR)*(x-1))/x });

\addplot[domain=0.1:2^(2*\RR)-1, color=orange, samples=37, style=thick, mark=*, mark options={scale=0.2}] 
({x},{
2^(-(4/\MM)*(\RR +(\MM-1)*ln(1+ x)/2/ln(2)))
+((1-2^(-2*\RR))*(1-2^(-2*\RR) + 2^(-(2/\MM)*(\RR-ln(1+ x)/2/ln(2))) ) )/ x
  });

\addplot[domain=0.1:2^(2*\RR)-1, color=blue, samples=37, style=thick, mark=|, mark options={scale=0.5}] 
({x},{1/(1+x) + x/(1 + x)*2^(-2*\RR)});

\addplot[domain=2^(2*\RR)-1:100, color=blue, samples=27, style=thick, mark=|, mark options={scale=0.5}] 
({x},{1/(1+x) + x/(1 + x)*2^(-2*\RR)});

\addplot[domain=2^(2*\RR)-1:100, color=red, samples=27, style=thick] 
({x},{1/(1+x) + 2^(-2*\RR)*x /(1+x) });

\addplot[domain=0.1:2^(2*\RR)-1, color=red, samples=37, style=thick] 
({x},{
1/(1+x) + x/(1+x) * 2^(-(2*\RR/\MM + (\MM-1)/\MM * ln(1+x)/ln(2))) });

\addplot[domain=0.1:100, color=black, samples=100, style=thick, dashed] 
(x,{1/(1+x)});

\addplot[domain=0.1:100, color=black, samples=100, style=thick, dotted] 
(x,{2^(-2*\RR)});

\node[] (Rcrit) at (axis cs: 15,0.4) {};


\addplot[dotted, dotted] 
coordinates {(2^(2*\RR)-1,0)  (2^(2*\RR)-1,0.3)};

\end{semilogxaxis}
        
\node at (Rcrit) {\scriptsize $2^{2R}-1$};        
\end{tikzpicture}

\caption{
Mean-squared error (MSE) in the Gaussian multiple observation setting \eqref{eq:AWGN} with compression. Here the number of measurement nodes is $m=3$. All measurements have the same SNR $\gamma/m$. The maximum likelihood (ML) RD risk $D^*_{\phi^{\ML}}(R)$ and the Bayes RD risk $D^*(R)$ correspond to compressing the noisy measurements using a good RD code of the CE strategy. The indirect DR function $D_{\ind}(R)$ corresponds to the optimal EC strategy. The minimal MSE (MMSE) corresponds to estimation without compression, the infinite bitrate asymptotic. Right: MSE as a function of the SNR parameter $\gamma$; we fix $R=2$. Left: MSE as a function of compression bitrate $R$; we fix $\gamma = 5$. }
\label{fig:Gaussian_centralized}
\end{center}
\end{figure*}

By comparing \eqref{eq:Gaussian_centralized} and \eqref{eq:idrf_Gaussian}, we conclude that the Bayes RD risk coincides with the indirect DR function either when $R$ is smaller than $R_{\gamma}$ or when $m=1$. In other words, in these cases, it is possible to attain the optimal performance even when the CE approach is utilized; knowledge of the observation model \eqref{eq:AWGN} is not required.
On the other hand, when the rate $R$ is larger than $R_{\gamma}$, the exponential convergence of the RD risk to $(1+\gamma)^{-1}$ is $m$ times slower than the equivalent convergence of the indirect DR function. The RD risk of the ML estimator $D^*(R)$ converges to its infinite bitrate asymptotic $1/\gamma$ at the same rate as the Bayes RD risk. 
A comparison between $D_{\phi^\ML}^*(R)$, $D^*(R)$, and $D_{\ind}(R)$ is given in Figure~\ref{fig:Gaussian_centralized}.

\section{Conclusions}
\label{sec:Conclusions}
We considered the problem of estimating a latent signal from a dataset undergoing lossy compression. We focused on a compress-and-estimate (CE) scenario: the data is compressed using a rate-distortion (RD) code attaining the Shannon RD function of the data with equality and the underlying signal is estimated from the compressed version of the data. While this approach is sub-optimal in general, it is utilized in many cases when the optimal compression approach to the problem at hand is infeasible. 
To characterize the CE performance, we defined the RD risk of an estimator as its risk evaluated on data sampled from the RD-achieving distribution. We showed that whenever this distribution satisfies a suitable transportation cost (TC) inequality and
the output distribution of the compression code and the RD-achieving distribution converge in relative entropy, the performance of any Lipschitz estimation procedure from the data is given by its corresponding RD risk. The TC condition holds, broadly speaking, whenever the model is discrete and separable, or when the RD-achieving distribution is the Gaussian distribution. The other necessary condition, convergence in relative entropy, is known to hold for various random coding constructions. 
\par
In order to illustrate the usefulness of this characterization and the differences between the CE and the optimal performance, we evaluated the RD risk in the case of a binary memoryless signal observed under a binary symmetric channel and a Gaussian signal observed through multiple AWGN channels. This evaluation reveals several interesting differences between CE and the optimal source coding performance. \par
\medskip
The work presented here leaves open various research challenges. 
Most notably, our main results were obtained by reducing the asymptotic equivalence of the CE and RD risks to convergence in relative entropy of the posterior distribution of a good RD code. Our most limiting conditions originate from the need to bound the relative entropy from above. We postulate that these assumptions can be relaxed by taking an approach that does not involve convergence in relative entropy. For example, it may be possible to characterize the distribution of good RD codes in terms of the Wasserstein distance without utilizing transportation-cost inequalities. %
Another line of future work is the extension of the results provided here to multi-terminal settings, i.e. when the data is compressed at multiple locations. %

\section*{ACKNOWLEDGMENT}
The authors are grateful for all the Reviewers and the Associate Editor for their insightful comments and suggestions that have greatly improve the clarity of the paper and its contributions. 

\appendices

\section{Derivation of Equation~\ref{eq:2d_diff}
\label{app:proof_intro}
}
Set $\alpha = \frac{\sqrt{\gamma/2}}{1+\gamma}$ and assume that the data is 
\[
U_i \equiv \ex{\theta_i|Y_i} = \alpha \left(Y_{1,i}+Y_{2,i}\right),\quad i=1,\ldots,n,
\]
which is iid Gaussian with variance $\frac{\gamma}{1+\gamma}$. Denote by $P_{U_1 \hat{U}^*}$ the single-letter RD-achieving distribution of $U^n$ under quadratic distortion. The forward channel representation of this distribution is
\[
\hat{U}^* \overset{D}{=} (1-2^{-2R})U_1 + \sqrt{2^{-2R} (1-2^{-2R}) \var[U_1] }W',
\]
for some $W'\sim \Ncal(0,1)$, independent of $U_1$. Since $\hat{U}^*$ and $Y_{1,1}$ are jointly Gaussian, the Bayes estimator of $Y_{1,1}$ from $\hat{U}^*$ is
\begin{align}
\nonumber
 \phi_1^B(\hat{U}^*) & = \frac{\ex{Y_{1,1} \hat{U}^* }}{\var[\hat{U}^*]}\hat{U}^* = \frac{(1-2^{-2R})\ex{Y_{1,1} U_1 }}{(1-2^{-2R})\var[U_1]} \hat{U}^*
\nonumber \\
& = \frac{\alpha}{\var[U_1]}(\var[Y_{1,1}] + \ex{Y_{1,1} Y_{1,2}} ) \hat{U}^* \nonumber  \\
& =  \frac{1}{\sqrt{2\gamma}}(1+\gamma/2 + \gamma/2)\hat{U}^* = \frac{1+\gamma}{\sqrt{2\gamma}}\hat{U}^*. 
\end{align}
The risk of $\phi^B(\hat{U}^*)$ is
\begin{align*}
 & \ex{\left(Y_1 - \frac{1+\gamma}{\sqrt{2\gamma}} \hat{U}^*\right)^2}  = \var[Y_{1,1}] - \frac{\left(\ex{Y_{1,1}\hat{U}^*}\right)^2}{\var[\hat{U}^*]} \\
&\qquad = 1 + \frac{\gamma}{2} - \frac{(1-2^{-2R})^2 \alpha^2 (1+\gamma)^2}{(1-2^{-2R})\frac{\gamma}{1+\gamma}} \\
&\qquad = 1 + \frac{\gamma}{2} - (1-2^{-2R}) \frac{1+\gamma}{2} = \frac{1}{2} + 2^{-2R} \frac{1+\gamma}{2}.
\end{align*}
The Bayes RD risk is obtained by summing the risk of $\phi_1^B(\hat{U}^*)$ and of $\phi_2^B(\hat{U}^*)$, which are identical terms due to the symmetry in $Y_{1,1}$ and $Y_{1,2}$. Overall, the Bayes RD risk is 
\begin{align}
    \label{eq:BayesRD_intro}
 D^*_{U}(R) = 1 + 2^{-2(R-R_{\gamma})}, 
\end{align}
where $R_{\gamma} = \frac{1}{2}\log(1+\gamma)$. For comparison, the quadratic DR function of $Y^n$ of \eqref{eq:2d_Gaussian_model} is
\begin{align}
    \label{eq:DyR}
D_y(R) = 2^{-[R-R_{\gamma}]^+} + 2^{-2(R-R_{\gamma})+[R-R_{\gamma}]^+}.
\end{align}
Equation \eqref{eq:2d_diff} is the difference between \eqref{eq:BayesRD_intro} and \eqref{eq:DyR}.

\section{Proof of Proposition~\ref{prop:Talagrand_ext}
\label{app:Talagrand}
}
Let $W^n \sim \Ncal({\bf 0},\Id_n)$, $V^n \sim P$, and 
$\tilde{W}^n \sim \Ncal(\mu^n,\Sigma_n)$. Write $\Sigma_n = \Um \Lambda \Um^\top$ with $\Um$ unitary and $\Lambda$ diagonal with nonzero diagonal elements. Also write $\Lambda = \Lambda^{1/2}\Lambda^{1/2}$. Using Proposition~\ref{prop:Talagrand}, 
 for any coupling of $W^n$ and $V^n$, 
 \begin{align}
 & 2 D(P \parallel  \Ncal({\bf 0},\Id_n)) \geq \left\|W^n - V^n \right\|_2^2 \nonumber \\
	&\quad  = \frac{\left\| \Lambda^{1/2} \right\|_2^2}{\left\| \Lambda^{1/2} \right\|_2^2} \left\|W^n-y^n - V^n-y^n\right\|_2^2 \nonumber \\
	&\quad  \geq 
	\frac{1}{\left\| \Lambda^{1/2} \right\|_2^2} \left\| \Lambda^{1/2} (W^n-\mu^n) - \Lambda^{1/2}(V^n-\mu^n)\right\|_2^2  \nonumber \\
	&\quad  = \frac{1}{\left\| \Lambda^{1/2} \right\|_2^2} \left\| \Um \Lambda^{1/2} (W^n-\mu^n) - \Um \Lambda^{1/2}(V^n-\mu^n)\right\|_2^2 \nonumber \\
	&\quad  = \frac{1}{\left\| \Lambda^{1/2} \right\|_2^2} \left\| \tilde{W}^n - \tilde{V}^n\right\|_2^2 = \frac{1}{\left\| \Lambda \right\|_2} \left\| \tilde{W}^n - \tilde{V}^n\right\|_2^2, \label{eq:Talagrand_proof}
	\end{align}
	where $\tilde{V}^n \sim \tilde{P}$,  with $\tilde{P}$ an affine invertible transformation of $P$. Since the relative entropy is invariant to such transformations, i.e.,
	$D(P\parallel \Ncal({\bf 0},\Id_n)) = D(\tilde{P} \parallel \Ncal(\mu^n,\Sigma_n))$,
	\eqref{eq:Talagrand_ext} follows by taking the infimum of \eqref{eq:Talagrand_proof} over all couplings of $\tilde{W}^n$ and $\tilde{V}^n$, dividing both sides by $n$, and taking their square roots. 

\section{\label{app:proof:main}Proof of Lemma~\ref{lem:main}}
For the sake of clarity, we omit the superscripts $n$ as they can be deduced from the context. We have
\begin{subequations}
\label{eq:main_proof}
\begin{align}
& \left| \left( D_\phi^\CE(\theta) \right)^{1/p} -  \left( D^*_\phi(R(D_y),\theta) \right)^{1/p}
    \right| \nonumber \\
& = \left| \left( \ex{\rho^p\left(\theta,\phi(Z^n)\right)}  \right)^{1/p} -  \left( \ex{\rho^p\left(\theta,\phi(Z^{*n})\right)}  \right)^{1/p}
    \right| 
\\
    & \qquad \overset{(a)}{\leq} L \cdot W_p(P_{Z|Y}, P_{Z^{*}|Y}|P_Y) \\
    &  \qquad  \overset{(b)}{=} L \cdot \int_{\Yc} W_p\left(P_{Z|Y=y},P_{Z^{*}|Y=y} \right) \diff P_Y(y) \\
& \qquad 
\overset{(c)}{\leq}
L \cdot \int_{\Yc}   \sqrt{  2c(y) D(P_{Z|Y=y} \parallel  P_{Z^{*}|Y=y})} \diff P_Y(y) \\
& \qquad  \overset{(d)}{\leq}
L \cdot  \sqrt{ \int_{\Yc} 2c(y) \diff P_Y(y)} \\
& \qquad \qquad \times  \sqrt{ \int_{\Yc}  D(P_{Z|Y=y} \parallel  P_{Z^{*}|Y=y}) \diff P_Y(y) }  \\
& \qquad \overset{(e)}{=} L \cdot \sqrt{ 2\ex{c(Y)} D\left( P_{UV} \parallel P_{U'V} \right) } , \label{eq:main_proof}
\end{align}
\end{subequations}
where (a) is due to Proposition~\ref{prop:Lipschitz}, 
(b) follows from the definition of the conditional Wasserstein distance in \eqref{eq:cond wass}, 
(c) follows because 
\[
W_p\left(P_{Z|Y=y},P_{Z^{*}|Y=y} \right) \leq \sqrt{ 2c(y) D(P_{Z|Y=y} \parallel  P_{Z^{*}|Y=y}) },
\]
holds for almost every $y$ by the lemma assumption, (d) follows from the Cauchy-Schwarz inequality, and (e) follows from the identity
\[
\int_{\Yc} D(P_{Z|Y=y} \parallel  P_{Z^{*}|Y=y}) \diff P_Y(y) = D\left( P_{UV} \parallel P_{U'V} \right). 
\]
The assumption $P_{Z|Y=y} \ll P_{Z^*|Y=y}$ (Assumption \ref{ass:A-C1}) implies that the support of $P_{Z|Y=y}$ is contained in the support of $P_{Z^*|Y=y}$, allowing us to write
\begin{align*}
    & D(P_{ZY} \parallel  P_{Y|Z^{*}} P_{Z}) = 
    D(P_{ZY} \parallel  P_{Z^{*}Y}) - D(P_Z\parallel P_{Z^{*}}). 
\end{align*}
Consequently, Proposition~\ref{prop:KostinaVerdu} leads to 
\begin{align}
  & D(P_{ZY} \parallel  P_{Z^*Y})  = D(P_{Y Z} \parallel  P_{Y|Z^{*}}P_Z) + D(P_Z \parallel  P_{Z^{*}}) \nonumber \\
& \qquad \qquad \leq I(Y^n ; Z^n) - \Rcal_{Y^n}(D_y) + D(P_Z \parallel  P_{Z^{*}}) \nonumber \\
& \qquad \qquad \leq n R_n - nR(D_y) + D(P_Z \parallel  P_{Z^{*}}) \label{eq:display}.
\end{align}
In \eqref{eq:display} we used that $nR(D_y) \leq \Rcal_{Y^n}(D_y)$ which follows from the definition of $R(D_y)$, and that $I(Y^n;Z^n) \leq n R_n$, which follows from the data processing inequality as the code is $D_y$-admissible. Combining \eqref{eq:display} with \eqref{eq:main_proof}, we obtain
\begin{align} 
    & \left| \left( D_\phi^\CE(\theta) \right)^{1/p} -  \left(D^*_\phi(R(D_y),\theta) \right)^{1/p}
    \right|  \nonumber \\
    & \leq 
     L \sqrt{2 \ex{c(Y^n)} \left( n R - n R(D_y) + D(P_{Z^n}\parallel P_{Z^{*n}}) \right)},
  \nonumber
\end{align}
which is equivalent to the claimed inequality \eqref{eq:main_2}. 

\section{Proof of Theorem~\ref{thm:main_product}
\label{proof:thm:main_product}
}

We show that the conditions of Lemma~\ref{lem:main} hold in this special case with $p=1$, $\Theta = \Theta_1^n$, $\theta = \theta^n$, $P_\theta = P_{\theta_1}^n$, $\nu = \rho_H$, and $\rho$ of Lemma~\ref{lem:main} is of the form
\[
\rho(\theta^n,\hat{\theta}^n) \equiv \frac{1}{n} \sum_{i=1}^n \rho(\theta_i,\hat{\theta}_i). 
\]
We also use $\psi = \phi^n$, where $\phi^n : \Zcal^n \to \Theta_1^n$ with its $i$-th coordinate $[\phi^n(z^n)]_i$ equals $\phi(z_i)$. Because condition (iii) of Section~\ref{subsec:discrete} says that $\rho$ is bounded by $M$, for any $z^n,\hat{z}^n \in \Zcal^n$ we have
\begin{align*}
    & \rho\left(\phi^n(z^n),\phi^n(\hat{z}^n) \right) = \frac{1}{n} \sum_{i=1}^n \rho(\phi(z_i),\phi(\hat{z}_i)) \\
    & = \frac{1}{n} \sum_{i\,:\,z_i\neq \hat{z}_i} M \leq M \rho_H(z^n,\hat{z}^n).
\end{align*}
Namely, $\phi^n$ is $M$-Lipschitz with respect to the normalized Hamming distance on $\Zcal^n$. For any $y^n \in \Ycal^n$, Proposition~\ref{prop:Marton} implies that $P_{Z^{* n}|Y^n=y^n}$ is $\TC(1/4n)$. Finally, since $\rho$ is bounded, the conditions
$\ex{\rho(\theta^n,\phi^n(Z^{* n}))} < \infty$ and 
$\ex{\rho(\theta^n,\phi^n(Z^{n}))} < \infty$ are satisfied. \par
Applying Lemma~\ref{lem:main}, we get
\begin{align*}
    & \left| \frac{1}{n} \sum_{i=1}^n \ex{ \rho \left(\theta_i, \phi(Z_i) \right) } - 
    \frac{1}{n} \sum_{i=1}^n \ex{\rho\left(\theta_i, \phi( Z^*_i )\right) } 
    \right| \\
    & \leq 
     \frac{M}{\sqrt{2}} \sqrt{ R_n - R(D_y) + \frac{1}{n} D(P_{Z^n}\parallel P_{Z^{*  n}}) }.
\end{align*}
From $R_n \to R(D_y)$ as $n\to \infty$ and Assumption \ref{ass:A-C2}, we obtain \eqref{eq:main_prod_1}. This complete the proof of (i). In part (ii), $\phi_n : \Zcal^n \to \Theta_1^n$ is an arbitrary $L$-Lipschitz function. 
First note that \eqref{eq:main_prod_2} trivially holds whenever $\ex{\rho(\theta_i, [\phi_n(Z^n)]_i)} = \infty$, for some $i,n \in \mathbb N$, where we use the notation $[v^n]_i \equiv v_i$. 
For the complementary case, we use Lemma~\ref{lem:main} with the setting as in part (i) except that $\psi = \phi_n$. We obtain:
\begin{align} \label{eq:main_prod_proof1}
     & \frac{1}{n} \sum_{i=1}^n \ex{\rho\left(\theta_i, [\phi_n( Z^{*  n})]_i \right) } \\
     & \qquad \leq 
     \frac{L}{\sqrt{2}} \sqrt{ R_n - R(D_y) + \frac{1}{n}D(P_{Z^n}\parallel P_{Z^{* n}})}  \nonumber \\
      &\qquad \qquad  + \frac{1}{n} \sum_{i=1}^n \left(\ex{ \rho \left(\theta_i, [\phi_n(Z^n)]_i \right) }\right). \nonumber 
\end{align}
Now, from  $P_{\theta^n Y^n Z^{*n}} = \prod_{i=1}^n P_{\theta_i Y_i Z^*_i}$, we get that $\ex{\rho\left(\theta_i, [\phi_n( Z^{*n})]_i \right)}$ is bounded from below by 
\[
\ex{\rho\left(\theta_1,\phi^B(Z_1^*) \right)} \equiv D^*(R(D_y)),
\]
where $\phi^B$ is the Bayes optimal estimator of $\theta_1$ from $Z_1^*$ of \eqref{eq:Bayes_estimator}. Hence
\begin{align}
    \label{eq:display2}
\frac{1}{n} \sum_{i=1}^n \ex{\rho\left(\theta_i, [\phi_n( Z^{*  n})]_i \right) }  \leq D^*(R(D_y)).
\end{align}
By assumption, we have that $R_n-R(D_y) \to 0$ and, due to Assumption \ref{ass:A-C2}, $D(P_{Z^n}\parallel P_{Z^{*  n}})/n\to 0$. Using these facts in \eqref{eq:main_prod_proof1} and combining \eqref{eq:display2}, we obtain \eqref{eq:main_prod_2}.

\section{Proof of Theorem~\ref{thm:main_Gaussian1}
\label{proof:thm:main_Gaussian1}
}
We first show that Lemma~\ref{lem:main} takes the following simplified form under the quadratic Gaussian setting.
\begin{lem} \label{lem:main_Gaussian}
Under the quadratic Gaussian setting, suppose that $P_{Z^n|Y^n} :\mathbb R^n \to \reals^n$ is a $D_y$-admissible code of rate $R_n$ such that $P_{Z^n|Y^n=y^n}$ has a density in $\reals^n$. For any $L$-Lipschitz $\phi : (\mathbb R^n,\frac{1}{\sqrt{n}}\|\cdot\|_2) \to (\Theta, \rho)$, 
\ea{
    & \left| \sqrt{ \ex{ \rho^2 \left(\theta, \phi(Z^n) \right) }} -
    \sqrt{D_{\phi}^*(R(D_y),\theta)}
    \right| \nonumber \\
    & \leq 
     L  \sqrt{2 \left\| \Sigma_n \right\| \left( R_n - R(D_y) + \frac{1}{n}D(P_{Z^n}\parallel P_{Z^{* n}}) \right) },
\label{eq:main_Gaussian_1}
}
provided $\ex{\rho^2 (\theta, \phi(Z^n))} < \infty$ and $D_{\phi}^*(R(D_y),\theta) < \infty$. 
\end{lem}
\subsubsection*{Proof of Lemma~\ref{lem:main_Gaussian} }
We show that the conditions of Lemma~\ref{lem:main} are met in this case. 
First note that $d=\|\cdot\|_2^2$ is a sub-additive distortion measure and that $d_{\min}=0$. 
In addition, Assumption \ref{ass:A-RD} holds for $0\leq D \leq \tr (\Sigma_n)$, since in this case
$P_{Y^n Z^{*n}}$ as defined by \eqref{eq:Gaussian_channel} is the unique solution to the minimization problem \eqref{eq:RDF_def} when $P_{Y^n} = \Ncal(\mu^n,\Sigma_n)$; see \cite{berger1971rate}. 
For any $y^n \in \mathbb R^n$, $P_{Z^*|Y^n=y^n}$ is a Gaussian measure on $\mathbb R^n$, and by 
Proposition~\ref{prop:Talagrand_ext} it is $\TC(n^{-1}\|\Sigma_n\|)$ with respect to the normalized Euclidean distance and $p=2$. Furthermore, Assumption \ref{ass:A-C1} holds because $P_{Z^n|Y^n=y^n}$ has a density. 
As the conditions of Lemma~\ref{lem:main} hold with $p=2$, Lemma~\ref{lem:main_Gaussian} follows from \eqref{eq:main_2}. \\

To prove Theorem~\ref{thm:main_Gaussian1}, we use that $R_n \to R(D_y)$,  $n^{-1}D(P_{Z^n}\parallel P_{Z^{* n}})\to 0$ (Assumption \ref{ass:A-C2}), absolute continuity of $P_{Z^n|Y^n=y^n}$ with respect to the Lebesgue measure (Assumption \ref{ass:A-C1}), and that the sequence of Lipschitz estimators $\{\phi_n\}_{n=1}^\infty$ has a uniformly bounded Lipschitz constant. The required convergence now follows from  Lemma~\ref{lem:main_Gaussian}.

\section{\label{app:proof_gaussian_single} Proof of Proposition~\ref{prop:single_vector} }
Each entry of the vector-values sequence $\Yv^n$ is distributed as $\Ncal(0, \Sigma_{\Yv_1})$, where
\[
\Sigma_{\Yv_1} = {\av \av^\top}+\Id_m,\quad \av = (\gamma_1,\ldots,\gamma_m).
\]
The distribution $P_{\Zv^*_1 \Yv_1}$ achieving the single-letter RD function of $\Yv^n$ is known to satisfy the backward channel \cite{berger1971rate}
\begin{equation} \label{eq:backward_channel}
 \Yv_1 = \Zv_1^* +  \Uv \Tv \Wv,
\end{equation}
where:
\begin{itemize}
\item[(i)] $\Um$ is a unitary matrix such that
\[
\Uv^\top \Sigma \Uv = \mathrm{diag}\left[\lambda_1(\Sigma_{\Yv_1}),\ldots,\lambda_L(\Sigma_{\Yv_1}) \right],
\]
$\lambda_1(\Sigma_{\Yv_1}) = 1+\gamma$, and $\lambda_i(\Sigma_{\Yv_1}) = 1$, for $i=2,\ldots,m$.

\item[(ii)] $\Wv$ is a standard normal vector  independent of $\Zv_1^*$. 

\item[(iii)] $\Tv$ is a diagonal matrix with $\Tv_{i,i} = \min \left\{\eta,\lambda_i(\Sigma_{\Yv_1}) \right\}$.
\item[(iv)] $\eta$ is a parameter satisfying
\ea{
R = \sum_{i=1}^m R_i \equiv \frac{1}{2m}\sum_{i=1}^m \log^+ \lb  \f {\lambda_i(\Sigma_{\Yv_1})}{\eta} \rb.
\label{eq:waterfilling_indirect}
}
\end{itemize}
The Bayes RD risk is the minimal MSE (MMSE) in estimating the vector $\Xv_1$ from $\Zv_1^*$. Define ${\mathbf \Zt}_1^*  \equiv \Uv^\top \Zv_1^*$, and note that 
\begin{align*}
D_y(R) & = \tr \left( \ex{ (\Yv_1-\Zv_1^*)^\top(\Yv_1-\Zv_1^*)} \right)\\
& = \tr  \left(\ex{ (\Uv^\top \Yv_1-{\mathbf \Zt}_1^*)^\top(\Uv^\top \Yv_1-{\mathbf \Zt}_1^*)} \right) \\
& = \sum_{i=1}^m D_i, 
\end{align*}
where $D_i \equiv \min\{ \lambda_i(\Sigma_{\Yv_1}), \eta \}$. 
In addition, note that $P_{{\mathbf \Zt}^*_1|\Yv_1}$ can be represented by the Gaussian channel
\ea{
\Zt_{1,i}^* & = \left(1-2^{-2R_i} \right) [\Uv^\top \Yv_1]_i + \sqrt{D_i(1-2^{-2R_i}) }W'_i,\nonumber \\
& = \left(1-2^{-2R_i} \right)\uv_i^\top \av \theta_1 + \left(1-2^{-2R_i} \right) [\uv_1^\top \Wv]_i  \nonumber \\
& \quad \quad + \sqrt{D_i(1-2^{-2R_i}) }W'_i  \nonumber \\
& = \left(1-2^{-2R_i} \right)\uv_i^\top \av \theta_1 \nonumber \\
& \quad \quad + \sqrt{D_i(1-2^{2R_i}) + (1-2^{-2R_i})^2 } W_i'',
\label{eq:scalar_relation}
}
where $(W'_1,\ldots,W'_m)$ and $(W''_1,\ldots,W_m'')$ denote mutually independent standard normal vectors that are independent of $\Wv$ and $\theta_1$. We write \eqref{eq:scalar_relation} in the matrix form 
\[
{\mathbf \Zt}_1^* = \Bv \theta_1 + \Cv \Wv''.
\]
The MMSE of $\theta_1$ given $\Zv_1^*$ has the form
\[
\mmse(\theta_1|\Zv_1^*) \equiv \var\left[\theta_1\right] - \var\left[\ex{\theta_1|\Zv_1^*}\right]. 
\]
Using the matrix inversion lemma, this MMSE satisfies \cite[Eq. 198]{wu2012optimal}:
\begin{align*}
& \mmse(\theta_1| \Zv_1^*) = \mmse({\theta_1|{\mathbf \Zt}_1^*}) \\
& \qquad \qquad = \left(1+ \Bv^\top \Sigma_{\Cv}^{-1} \Bv \right)^{-1} \\
&  \qquad \qquad = \left(1+ \sum_{i=1}^m \frac{(1-2^{-2R_i})( \av^\top \uv_i)^2 } {(1-2^{-2R_i})+D_i } \right)^{-1} \\
&  \qquad \qquad =  \left( 1+ \sum_{i=1}^m \frac{(1-2^{-2R_i})( \av^\top \uv_i)^2 } {(1-2^{-2R_i})+\min\left\{\eta,\lambda_i \right\} } \right)^{-1}.
\end{align*}
Since $\mathbf u_1 = \av/\|\av\|_2$, $R_1 = \frac{1}{2} \log \left( (1+\gamma)/\eta\right)$ and $\uv_2,\ldots,\uv_m$ are orthogonal to $\av$, we get
\begin{align*}
\mmse\left(\theta_1|{\mathbf \Zt}_1^*\right) & = \left( 1+ \gamma \frac{ 1+\gamma-\eta  } {1 + \gamma(\eta+1) } \right)^{-1}  \\
& =  \frac{1}{\gamma+1} + \frac{\gamma}{(\gamma + 1)^2} \eta.
\end{align*}
Equation \eqref{eq:Gaussian_centralized}
is obtained by setting $\eta' = \eta/(1+\gamma)$.

\section{Proof of Proposition~\ref{prop:ML}
\label{app:proof_Gaussian_ML}
}
Let us adopt the same notation as in the proof of Proposition~\ref{prop:single_vector}. From ${\mathbf \Zt}_1 = \Uv^* \Zv_1$, we have
\begin{align*}
\Zv_1 & = \left( \sum_{i=1}^m (1-2^{-2R_i}) \uv_i \uv_i^\top \right)\Yv_1 \nonumber \\
& \quad \quad + 
\left( \sum_{i=1}^m \sqrt{D_i (1-2^{-2R_i})} \uv_i \uv_i^\top \right)\Wv. 
\end{align*}
When $R < R_{\gamma}= \frac{1}{2}\log(1+ \gamma)$, 
we have
$\eta \geq 1$, $R_1=R$, $D_1 = (1+\gamma)2^{-2R}$, while $R_i =0$ and $D_i = 1$, for $i = 2,\ldots,m$. When $R \geq R_{\gamma}$, we have $\eta = 2^{-2(R-\frac{1}{2}\log(1+ \gamma))/m}$, $R_1 = \frac{1}{2}\log\frac{1+\gamma}{\eta}$, $R_i = \frac{1}{2}\log\frac{1}{\eta}$, $i=2,\ldots,m$, and 
$D_i = \eta$ for all $i=1,\ldots,m$. 
The ML estimator of $\theta_1$ from $\Yv_1$ is 
\[
\phi^{\ML}(\yv) \equiv  \frac{\av^\top \yv}{\gamma}. 
\]
Because $\uv_1 = \av / \|\av\|_2$ and $\{\uv_i\}_{i=2,\ldots,m}$ are orthogonal to $\av$, we get
\begin{align*}
&\medmath{ \phi^{\ML}(\Zv_1) = 
 (1-2^{-2R_1}) \frac{\av^\top}{\|\av\|_2^2} \Yv_1 + 
\left(  \sqrt{D_1 (1-2^{-2R_1})}\frac{\av^\top}{\|\av\|_2^2} \right)\Wv} \\
& \medmath{=  (1-2^{-2R_1}) \frac{\av^\top}{\|\av\|_2^2}  (\av X + \Wv') + 
\left( \sqrt{D_1 (1-2^{-2R_1})}\frac{\av^\top}{\|\av\|_2^2} \right)\Wv.}
\end{align*}
When $R < R_{\gamma}= \frac{1}{2}\log(1+ \gamma)$, 
we have
$\eta \geq 1$, $R_1=R$, $D_1 = (1+\gamma)2^{-2R}$, while $R_i =0$ and $D_i = 1$, for $i = 2,\ldots,m$. When $R \geq R_{\gamma}$, we have $\eta = 2^{-2(R-r_\gamma)/m}$, $R_1 = \frac{1}{2}\log\frac{1+\gamma}{\eta}$, $R_i = \frac{1}{2}\log\frac{1}{\eta}$, $i=2,\ldots,m$, and 
$D_i = \eta$ for all $i=1,\ldots,m$. 
The following equality in distribution holds:
\begin{align*}
    & \theta_1 - \phi^{\ML}(\Zv_1) \overset{D}{=} \\
    & 2^{-2R_1} \theta_1
     + \sqrt{ \frac{(1-2^{-2R_1})^2 + D_1 (1-2^{-2R_1}) }{\gamma }}W''
     \\
     & = \medmath{ \begin{cases}
       2^{-2R} \theta_1 - \sqrt{ \frac{(1-2^{-2R})^2 + (1+\gamma)2^{-2R}(1-2^{-2R}) }{\gamma }} W''
       &  R \leq R_{\gamma} \\
        \frac{\eta}{1+\gamma} \theta_1
         -  
        \sqrt{ \frac{(1-2^{-2R})^2 + \eta (1-2^{-2R}) }{\gamma }} W''
       & R > R_{\gamma}.
     \end{cases} 
     }
\end{align*}
The RD risk of $\phi^{\ML}$ is the variance of the last expression. Substituting $\eta = 2^{-\frac{2}{m}(R-R_\gamma)}$ for $R > R_{\gamma}$, we obtain 
\begin{align*}
& D^*_{\phi^{\ML}}(R) \\
& = \medmath{
\begin{cases}
       2^{-4R} + \frac{1}{\gamma }(1-2^{-2R})(1 + \gamma2^{-2R})
       &  R \leq R_{\gamma} \\
         \frac{\eta^2}{(1+\gamma)^2}
         +
 \frac{1}{\gamma}\left((1-2^{-2R})^2 + \eta (1-2^{-2R})\right)
       & R > R_{\gamma}
     \end{cases} } \\
    = & \medmath{ 
     \begin{cases}
        \frac{1}{\gamma}\left(1+2^{-2R}(\gamma-1)\right)
       &  R \leq R_{\gamma} \\
       \begin{multlined}[b]
       \left(
         \frac{2^{-2R}}{
         (1+\gamma)^{m-1}
         } \right)^{\frac{2}{m}} 
         +
        \frac{1}{\gamma}\left(1-2^{-2R}\right )\\
        \qquad 
        \times \left(1-2^{-2R} +\left(2^{-2R} (1+\gamma) \right)^{\frac{1}{m}} \right)
        \end{multlined}
       & R > R_{\gamma}.
     \end{cases}
     }
\end{align*}
The last display is identical to \eqref{eq:ML}.

\bibliographystyle{IEEEtran}
\bibliography{IEEEfull,compressAndEstimate}

\end{document}